\documentclass[authoryear]{elsarticle}

\usepackage{color,hyperref}
\journal{NeuroImage}

\newcommand{\mc}{\mathcal}
\newcommand{\real}{\mathbb{R}}

\newcommand{\map}[3]{#1: #2 \rightarrow #3}

\newcommand{\A}{\mathbf{A}}
\newcommand{\B}{\mathbf{B}}
\newcommand{\E}{\mathbf{E}}

\newcommand{\I}{\mathbf{I}}
\newcommand{\bvec}[2]{\begin{bmatrix} #1 \\ #2 \end{bmatrix}}
\newcommand{\bmat}[4]{\begin{bmatrix} #1 & #2 \\ #3 & #4 \end{bmatrix}}

\newcommand{\bb}{\mathbf{b}}
\newcommand{\bc}{\mathbf{c}}
\newcommand{\bx}{\mathbf{x}}
\newcommand{\bu}{\mathbf{u}}
\newcommand{\bp}{\mathbf{p}}

\newcommand{\rev}[1]{\textcolor{black}{#1}}

\usepackage{amsmath}
\usepackage{amsfonts}
\usepackage{amssymb}
\usepackage{graphicx}
\usepackage[margin=0.5in]{geometry}

\begin{document}
\begin{frontmatter}
	
\title{Optimal Trajectories of Brain State Transitions}
	
	\author[adr1,adr2]{Shi Gu}
	\author[adr2]{Richard F. Betzel}
	\author[adr3]{Marcelo G. Mattar}
	\author[adr4]{Matthew Cieslak}
	\author[adr4,adr5]{Philip R. Delio}
	\author[adr4]{Scott T. Grafton}
	\author[adr6]{Fabio Pasqualetti}
	\author[adr2,adr7]{Danielle S. Bassett\corref{mycorrespondingauthor}}
	\cortext[mycorrespondingauthor]{Corresponding author}
	\ead{dsb@seas.upenn.com}

	\address[adr1]{Applied Mathematics and Computational Science, \\University of Pennsylvania, Philadelphia, PA, 19104 USA}
	\address[adr2]{Department of Bioengineering, \\University of Pennsylvania, Philadelphia, PA, 19104 USA}
	\address[adr3]{Princeton Neuroscience Institute, Princeton University, Princeton NJ 08544}
	\address[adr4]{Department of Psychological and Brain Sciences, \\University of California, Santa Barbara, CA, 93106 USA}
	\address[adr5]{Neurology Associates of Santa Barbara, \\Santa Barbara, CA, 93105 USA}
	\address[adr6]{Department of Mechanical Engineering, \\University of California, Riverside, CA, 92521 USA}
	\address[adr7]{ Department of Electrical \& Systems Engineering, \\University of Pennsylvania, Philadelphia, PA, 19104 USA}

	\begin{abstract}
 The complexity of neural dynamics stems in part from the complexity of the underlying anatomy. Yet how white matter structure constrains how the brain transitions from one cognitive state to another remains unknown. Here we address this question by drawing on recent advances in network control theory to model the underlying mechanisms of brain state transitions as elicited by the collective control of region sets. We find that previously identified attention and executive control systems are poised to affect a broad array of state transitions that cannot easily be classified by traditional engineering-based notions of control. This theoretical versatility comes with a vulnerability to injury. In patients with mild traumatic brain injury, we observe a loss of specificity in putative control processes, suggesting greater susceptibility to neurophysiological noise. These results offer fundamental insights into the mechanisms driving brain state transitions in healthy cognition and their alteration following injury.
	\end{abstract}
	
	\begin{keyword}
network neuroscience \sep control theory \sep traumatic brain injury \sep cognitive control \sep diffusion imaging
	\end{keyword}
	
\end{frontmatter}



\section*{Introduction}

The human brain is a complex dynamical system that transitions smoothly and continuously through states that directly support cognitive function \citep{deco2011emerging}. Intuitively, these trajectories can map out the mental states that our brain may pass through as we go about the activities of daily living. In a mathematical sense, these transitions can be thought of as trajectories through an underlying state space \citep{shenoy2011dynamical,freeman1994characterization,gu2016energy}. While an understanding of these trajectories is critical for our understanding of cognition and its alteration following brain injury, fundamental and therefore generalizable mechanisms explaining how the brain moves through states have remained elusive.

One key challenge hampering progress is the complexity of these trajectories, which stems in part from the architectural complexity of the underlying anatomy \citep{Hermundstad2011,Hermundstad2013, hermundstad2014structurally}. Different components (neurons, cortical columns, brain areas) are linked with one another in complex spatial patterns that enable diverse neural functions \citep{Rajan2016,Fiete2010,Levy2001}. These structural interactions can be represented as a graph or network, where component parts form the nodes of the network, and where anatomical links form the edges between nodes \citep{bullmore2009complex}. The architecture of these networks displays heterogenous features that play a role in neural function \citep{Medaglia2015},  development \citep{di2014unraveling}, disease \citep{braun2015human}, and sensitivity to rehabilitation \citep{weiss2011functional}. Despite these recent discoveries, how architectural features constrain neural dynamics in any of these phenomena is far from understood.

One simple and intuitive way to formulate questions about how neural dynamics are constrained by brain network architecture is to define a state of the brain by the $1 \times N$ vector representing magnitudes of neural activity across $N$ brain regions, and to further define brain network architecture by the $N \times N$ adjacency matrix representing the number of white matter streamlines linking brain regions \citep{gu2015controllability}. Building on these two definitions, we can ask how the organization of the white matter architecture constrains the possible states in which the brain can or does exist \citep{Durstewitz2008,Hansen2015}. Moreover, building on decades of cognitive neuroscience research that have carefully delineated the role of regional activation in cognitive functions \citep{Gazzaniga2013,szameitat2011test,alavash2015persistency}, we can then map brain states to cognitive processes, and extend our question to: how does the organization of white matter architecture constrain cognitive states \citep{Hermundstad2013,Hermundstad2014}, and the processes that enable us to move between those cognitive states \citep{cocchi2013dynamic}?
\begin{figure}
\centering
\includegraphics[width=0.7\linewidth]{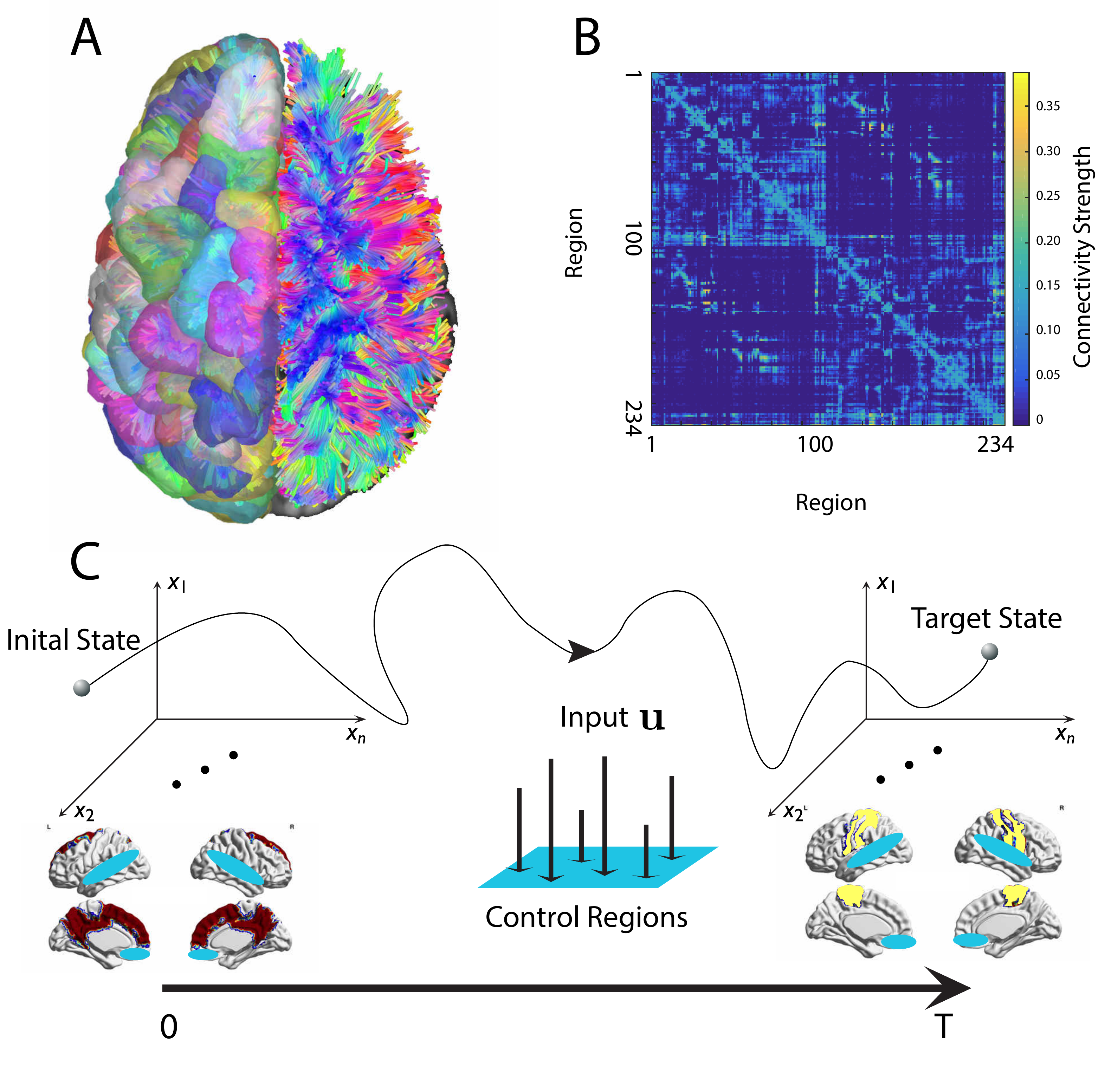}
\caption{{\bf Conceptual Schematic.} \emph{(A)} Diffusion imaging data can be used to estimate connectivity from one voxel to any other voxel via diffusion tractography algorithms. \emph{(B)} From the tractography, we construct a weighted network in which $N=234$ brain regions are connected by the quantitative anisotropy along the tracts linking them (see Methods).  \emph{(C)} We study the optimal control problem in which the brain starts from an initial state (red) at time $t = 0$ and uses multi-point control (control of multiple regions; blue) to arrive at a target state (yellow) at time $t = T$. }
\label{fig:figure1}
\end{figure}


To address these questions, we draw on recent advances in network control theory \citep{pasqualetti2014controllability} to develop a biologically-informed mathematical model of brain dynamics from which we can infer how the topology of white matter architecture constrains how the brain may affect (or \emph{control}) transitions between brain states. Within this model, we examine finite-time transitions (from initial to target state) that are elicited via the collective control of many regions, consistent with the collective dynamics observed to support cognition \citep{Salvador2005,Meunier2009,Power2011,Yeo2011} and action \citep{Bassett2011,Bassett2013,Bassett2015}. A natural choice for an initial state is the brain's well-known baseline condition, a state characterized by high activity in the precuneus, posterior cingulate, medial and lateral temporal, and superior frontal cortex \citep{Raichle2015,Raichle2007,Raichle2001}. While potential transitions from this \emph{default mode} are myriad, we focus this first study on examining transitions into target states of high activity in sensorimotor cortex: specifically the extended visual, auditory, and motor cortices. These states represent the simplest and most fundamental targets to transition from the default mode: for example, transitioning from the default mode to visual states might represent an immediate response to a surprising stimuli. Similarly, the transition from the default mode to motor states might represent the simple transition from rest to action.  Moreover, these transitions are of particular interest in many clinical disorders including stroke \citep{Carter2012} and traumatic brain injury \citep{Nudo2006,Lee2011} where the cognitive functions performed by these target areas are often altered, significantly effecting quality of life \citep{Kalpinski2013}.

Using network control theory, we examine the optimal trajectories from an initial state (composed of high activity in the default mode system) to target states (composed of high activity in sensorimotor systems) with finite time and limited energy. In this optimal control context, we investigate the role of white matter connectivity between brain regions in constraining dynamic state transitions by asking three interrelated questions. First, we ask which brain regions are theoretically predicted to be most energetically efficient in eliciting state transitions. Second, we ask whether these state transitions are best elicited by one of three well-known control strategies commonly utilized in mechanical systems \citep{gu2015controllability}. Third, we ask how specific each region's role is in these state transitions, and we compare this specificity between a group of healthy adults and a group of patients with mild traumatic brain injury.  In particular, the inclusion of this clinical cohort enables us to determine whether widespread injury leads to a decrement in the healthy network control profiles, thus requiring greater energy for the same functions, or an enhancement of the healthy network control profiles at the cost of a more fragile system, overly sensitive to external perturbations. Together, these studies offer initial insights into how structural network characteristics constrain transitions between brain states, and predict their alteration following brain injury.

To address these questions, we build structural brain networks from diffusion spectrum imaging (DSI) data acquired from 48 healthy adults and 11 individuals with mild traumatic brain injury (Fig.~1A). We perform diffusion tractography on these images to estimate the quantitative anisotropy along the streamlines linking $N=234$ large-scale cortical and subcortical regions extracted from the Lausanne atlas \citep{Cammoun2012,Daducci2012}. We summarize these estimates in a weighted adjacency matrix whose entries reflect the number of streamlines connecting different regions (Fig.~1B). We then define a model of brain state dynamics informed by the weighted adjacency matrix, and we use this model
to perform a systematic study of the controllability of the system. This construction enables us to examine how structural network differences between brain regions impact their putative roles in controlling transitions between cognitive states (Fig.~1C).

\section*{Materials and Methods}
\subsection*{Data Acquisition and Brain Network Construction}

Diffusion spectrum images (DSI) were acquired from 59 human adults with 72 scans in total, among which 61 scans were acquired from 48 healthy subjects (mean age $22.6\pm5.1$ years, 24 female, 2 left handed) and 11 were acquired from individuals with mild traumatic brain injury \citep{cieslak2014local}(mean age $33.8\pm13.3$ years, 4 female, handedness unclear). All participants volunteered with informed written consent in accordance with the Institutional Review Board/Human Subjects Committee, University of California, Santa Barbara. Deterministic fiber tracking using a modified FACT algorithm was performed until $100,000$ streamlines were reconstructed for each individual. Consistent with previous work
\citep{bassett2010efficient,bassett2011conserved,hermundstad2013structural,hermundstad2014structurally,klimm2014resolving, gu2015controllability,muldoon2016small,muldoon2016stimulation,sizemore2015classification},
we defined structural brain networks from the streamlines linking $N=234$ large-scale cortical and subcortical regions extracted from the Lausanne atlas \citep{hagmann2008mapping}. We summarize these estimates in a weighted adjacency matrix $\mathbf{A}$ whose entries $A_{ij}$ reflect the structural connectivity (quantitative anisotropy) between region $i$ and region $j$ (Fig.~1A). See SI for further details.

\subsection*{Network Control Theory}

Next, we consider the general question of how the brain moves between different states, where a state is defined as a pattern of activity across brain regions or voxels. In particular, we are interested in studying how the activity in individual brain regions affects the trajectory of the brain as it transitions between states; here, we define a trajectory as a set of states ordered in time. To address this question, we follow \citep{gu2015controllability,muldoon2016small,betzel2016optimally} by adopting notions from the emerging field of \emph{network control theory}, which offers a theoretical framework for describing the role of network nodes in the control of a dynamical networked system.

Network control theory is predicated on the choice of both a structural network representation for the system, and a prescribed model of node dynamics. In the context of the human brain, a natural choice for the structural network representation is the graph on $N$ brain regions whose $ij^{th}$ edge represents the QA between node $i$ and node $j$. The choice for the model of node dynamics is perhaps less constrained, as many models are available to the investigator.  These models range in complexity from simple linear models of neural dynamics with few parameters to nonlinear neural mass models with hundreds of parameters \citep{gu2015controllability,muldoon2016small}.

In choosing a model of neural dynamics to employ, we consider multiple factors. First, although the evolution of neural activity acts as a collection of nonlinear dynamic processes, prior studies have demonstrated the possibility of predicting a significant amount of variance in neural dynamics as measured by fMRI through simplified linear models
\citep{galan2008network,honey2009predicting,gu2015controllability}.
On the basis of this literature, we employ a simplified noise-free linear continuous-time and time-invariant network model
\begin{equation}
\label{eqn:ldn}
\dot{\bx}(t) = \A \bx(t) + \B \bu(t),
\end{equation}

\noindent where $\map{\bx}{\mathbb{R}_{\geq 0}}{\mathbb{R}^N}$ describes the state of brain regions over time, and $\A\in\real^{N\times N}$ is a symmetric and weighted adjacency matrix. The diagonal elements of the matrix $\A$ satisfy $A_{ii}=0$.
The input matrix $\B_{\mc{K}}$ identifies the control nodes $\mc{K}$ in the brain, where $\mc{K} = \{ k_1, \dots, k_m \}$ and
\begin{equation}
\B_{\mc{K}} = [e_{k_1}, \cdots, e_{k_2}]
\end{equation}
and $e_i$ denotes the $i$-th canonical vector of dimension $N$. The input $\map{\mathbf{u}_{\mc{K}}}{\real_{\geq
		0}}{\real^m}$
denotes the control strategy. Intuitively, this model enables us to frame questions related to brain state trajectories in a formal mathematics. Moreover, it allows us to capitalize on recent advances in network control theory \citep{pasqualetti2014controllability} to inform our understanding of internal cognitive control \citep{gu2015controllability,betzel2016optimally} and to inform the development of optimal external neuromodulation using brain stimulation \citep{muldoon2016stimulation}.

\subsection*{Optimal Control Trajectories}

Given the above-defined model of neural dynamics, as well as the structural network representation extracted from diffusion imaging data, we can now formally address the question of how the activity in individual brain regions affects the trajectory of the brain as it transitions between states.

We begin by defining an optimization problem to identify the trajectory between a specified pair of brain states that minimizes a given cost function. We define a cost function by the weighted sum of the energy cost of the transition and the integrated squared distance between the transition states and the target state. We choose this dual-term cost function for two reasons. First, theoretically, the energy cost term constrains the range of the time-dependent control energy $\mathbf{u}(t)$. In practice, this means that the brain cannot use an infinite amount of energy to perform the task (i.e., elicit the state transition), a constraint that is consistent with the natural energetic restrictions implicit in the nature of all biological systems but particularly neural systems \citep{Niven2008,Laughlin1998,Attwell2001,Laughlin2001}. Second, the term of the integrated distance term provides a direct constraint on the trajectory. Mathematically, this constraint penalizes trajectories that traverse states that are far away from the target state, based on the intuition that optimal transitions between states should possess reasonable lengths rather than being characterized by a random walk in state space. Together, these two terms in the cost function enable us to define an optimal control model from which we expect to find trajectories (from a given initial state to a specified target state) characterized by a balance between energy cost and trajectory length.

In the context of the optimization problem defined above, we wish to determine the trajectory from an initial state $\bx_0$ to a target state $\bx_T$. To do so, it suffices to solve the variational problem with the constraints from Equation \ref{eqn:ldn} and the boundary conditions for $\bx(t)$, i.e. $\bx(0)$ is the initial state and $\bx(T)$ is the target state. Note that here, the variational problem does not refer to the Bayesian variational inference, which tries to approximate an intractable posterior distribution. Instead, we use the term in the more traditional sense, and address the variational problem to infer a control input function $\mathbf{u(t)}$ to minimize the cost functional defined in Equation [4] with the boundary constraints. Mathematically, the variational problem is formulated as
\begin{equation}
\label{eqn_opt1}
\begin{aligned}
\min_\bu & & \int_{0}^T\left((\bx_T-\bx(t))^T(\bx_T-\bx(t)) + \rho \bu(t)^T \bu(t)\right) dt,\\
s.t. & & \dot{\bx}(t) = \A \bx(t) + \B \bu(t),\\
& & \bx(0) = \bx_0,\\
& & \bx(T) = \bx_T,
\end{aligned}
\end{equation}
where $T$ is the control horizon, $\rho\in\real_{> 0}$, and $(x_T - x(t))$ is the distance between the state at time $t$ and the target state.

To compute an optimal control $\bu^*$ that induces a transition from the initial state $\bx_0$ to the target state $\bx_T$, we define the Hamiltonian as
\begin{equation}
H(\bp,\bx,\bu,t) = \bx^T \bx + \rho \bu^T \bu+\bp^T(\A \bx+\B \bu).
\end{equation}
From the Pontryagin minimum principle \citep{boltyanskii1960theory}, if $\bu^*$ is an optimal solution to the minimization problem
with
corresponding state trajectory $\bx^*$, then there exists $\bp^*$ such that
\begin{eqnarray}
\frac{\partial{H}}{\partial{\bx}} & = & -2(\bx_T-\bx^*) + \A^T\bp^* = -\dot{\bp}^*,\\
\frac{\partial{H}}{\partial{\bu}} & = & 2\rho \bu^* + \B^T \bp^* = 0.
\end{eqnarray}
which reduces to
\begin{equation}
\label{eqn_1}
\bvec{ \dot{\bx}^*} {\dot{\bp}^* } = \begin{bmatrix} \A & -(2\rho)^{-1}\B\B^T\\ -2\I & -\A^T  \end{bmatrix}  \begin{bmatrix} {\bx}^*\\ {\bp}^*
\end{bmatrix}  +
\begin{bmatrix} \mathbf{0} \\ \I \end{bmatrix} 2\bx_T
\end{equation}
Next, we denote
\begin{eqnarray}
\tilde{\A} &=& \bmat{\A}{ -(2\rho)^{-1}\B\B^T}{-2\I}{-\A^T}, \\
\tilde{\bx}  & =&\bvec{\bx^*}{\bp^*}, \\
\tilde{\bb} & = &\bvec{\mathbf{0}}{\I}2\bx_T,
\end{eqnarray}
then Eqn [\ref{eqn_1}] can be written as
\begin{equation}
\dot{\tilde{\bx}} = \tilde{\A}\tilde{x} + \tilde{\bb},
\end{equation}
from which we can derive that
\begin{equation}
\label{eqn:final}
\tilde{\bx} + \tilde{\A}^{-1}\tilde{\bb} = e^{\A t} \tilde{\bc},
\end{equation}
where $\tilde{\bc}$ is a constant to be fixed from the boundary conditions.
Let $\tilde{\tilde{\bb}} = \bvec{\tilde{\tilde{\bb_1}}}{\tilde{\tilde{\bb_2}}} = \tilde{\A}^{-1}\tilde{\bb} $, $e^{-\A T} =
\bmat{\E_{11}}{\E_{12}}{\E_{21}}{\E_{22}}$ and plug in $t = 0, T$ with the corresponding $\bx_0$ and $\bx_T$, we have
\begin{eqnarray}
\label{eqn_l1}
\bvec{\bx(0)}{\bp(0)} + \bvec{\tilde{\tilde{\bb_1}}}{\tilde{\tilde{\bb_2}}} & = & \bvec{\tilde{\bc_1}}{\tilde{\bc_2}}, \\
\label{eqn_l2}
\bvec{\bx(T)}{\bp(T)} + \bvec{\tilde{\tilde{\bb_1}}}{\tilde{\tilde{\bb_2}}} & = & \bmat{\E_{11}}{\E_{12}}{\E_{21}}{\E_{22}}^{-1}
\bvec{\tilde{\bc_1}}{\tilde{\bc_2}}.
\end{eqnarray}
Note that from Equation[\ref{eqn_l1}], we can solve for $\tilde{\bc}_1$, where
\begin{equation}
\tilde{\bc_1} = \bx(0) + \tilde{\tilde{\bb}}_1.
\end{equation}
Finally, with $\tilde{\bc}_1$ on hand from Eqn[\ref{eqn_l2}], we can compute $\bp(T)$, where
\begin{equation}
\label{eqn_pT}
\bp(T) = \E_{12}^{-1}(\tilde{\bc}_1 -\E_{11}\tilde{\tilde{\bb}}_1-\E_{12}\tilde{\tilde{\bb_2}}-\E_{11}\bx(T)),
\end{equation}
with which we can finally get $\tilde{\bc}_2$, where
\begin{equation}
\label{eqn_c}
\tilde{\bc}_2 = \E_{21}\bx(T) + \E_{22} \bp(T)+ \E_{21}\tilde{\tilde{\bb_1}} +\E_{22}\tilde{\tilde{\bb_2}}
\end{equation}
and further the $\bu(t)$ and $\bx(t)$ from Equation \ref{eqn:final}.

Note that the formulae we derive here are the closed form solutions to the optimization objective, and therefore a numerical solver is not needed.

\subsection*{Statistics of Optimal Control Trajectories}

After calculating the optimal trajectories between initial and final states, we next sought to address the question of whether these trajectories differed in their energetic  and spatial requirements for different choices of control strategies, and between individual's whose brains were healthy and normally functioning, and individuals who had experienced a mild traumatic brain injury and had presented with complaints of mild cognitive impairment. To address this question, we
computed the energy cost of a trajectory, integrated over time $T$, as
\begin{equation}
E(\mathcal{K},\bx_0, \bx_T) = \int_{0}^{T}\bu_{\mathcal{K},\bx_0, \bx_T}^2 dt,
\end{equation}
and the spatial cost of a trajectory, integrated over time $T$, as
\begin{equation}
S(\mathcal{K},\bx_0, \bx_T) = \int_{0}^{T}\bx_{\mathcal{K},\bx_0, \bx_T}^2 dt,
\end{equation}
where $\bu_{\mathcal{K},\bx_0, \bx_T}$ is the associated control input and $\bx_{\mathcal{K},\bx_0, \bx_T}$ is the controlled trajectory with the given control set $\mc{K}$, initial state $\bx_0$ and the target
state $\bx_T$. We treat this energy as a simple statistic that can be compared across trajectories and subject groups, as an indirect measure from which we may infer optimality of cognitive function.

\subsection*{Control Efficiency}

The control efficiency is defined for each region to quantify its efficiency in affecting the transition from the default mode state to the three target states. Mathematically, suppose we have $N$ randomly chosen control sets, each indexed by $\mathcal{K}_1,\dots, \mathcal{K}_N$, for the target states $\bx _T^j$, $j= 1,2,3$, we calculate the corresponding optimal trajectory with respect
to $\mathcal{K}_k$ and denote the energy cost of the trajectory as $E(\mathcal{K}_k,\bx_0,\bx_T^j)$. The tiered value of the control set $\mathcal{K}_k$ for target $\bx_T^j$ is then defined as
\begin{equation}
t_{kj} = \sum_{l=1}^N \mathbf{1}(E(\mathcal{K}_l,\bx_0,\bx_T^j) > E(\mathcal{K}_k,\bx_0,\bx_T^j))
\end{equation}
where lower energy costs imply higher tiered values. The control efficiency for node $i$ in task $j$ is then
\begin{equation}
\mathrm{\zeta}_{ij} = \frac{\sum_{k=1}^N \mathbf{1}(i\in \mathcal{K}_k) \cdot t_{kj}}{\sum_{k=1}^N \mathbf{1}(i\in \mathcal{K}_k)}.
\label{eqn:CPS}
\end{equation}
or intuitively, the average of these tiered values.

\subsection*{Network Communicability to the Target State}

For a given weighted network $\mathbf{A}$, the network communicability $\mathbf{G}$ quantifies the extent of indirect connectivity among nodes. Here we adopt the generalized definition in \citep{crofts2009weighted} and define the network communicability as $\mathbf{G} =
\exp(\mathbf{D}^{-1/2}\mathbf{A}\mathbf{D}^{-1/2})$, where $\mathbf{D}$ is the diagonal matrix with the diagonal element ${D}_{ii}
= \sum_{i}A_{ij}$. For a given target state $\bx_T$, denote the set of active regions as $\mathcal{I}_{\bx_T}$, the communicability to the target states ($\mathrm{GT}_{i}$) is
then defined as the sum of communicability to all of the target regions, i.e. $\mathrm{GT}_{i} = \sum_{j\in \mathcal{I}_{\bx_T}} G_{ij}$. Further, the normalized network communicability to the target regions ($\mathcal{C}_i$) is then defined as
\begin{equation}
\mathcal{C}_i = \frac{\mathrm{GT}_i}{\sum_j\mathrm{GT}_j}.
\label{eqn:ngt}
\end{equation}
All results reported in this study are based on the normalized network communicability.

\subsection*{Energetic Impact of Brain Regions on Control Trajectories}
To quantify the robustness of controllability of a node when it is removed from the control set consisting of all nodes, we iteratively remove nodes from the network and compute the \emph{energetic impact} of each region on the optimal trajectory as the resulting increase in the log value of the energy cost. Intuitively, regions with high energetic impact are those whose removal from the network causes the greatest increase in the energy required for the state transition. Mathematically, denote $\mathcal{K}_0$ as the control set of all nodes and $\mathcal{K}_i$ as the control set without node $\mathcal{[K]}_i$, the energetic impact of node $i$ for target $\bx_T^j$ is defined as
\begin{equation}
\mathcal{I}_{ij} = \log\frac{E(\mathcal{K}_i,\bx_0,\bx_T^j)}{E(\mathcal{K}_0,\bx_0,\bx_T^j)}
\label{eqn:rc}
\end{equation}
which intuitively measures robustness controllability.
\section*{Results}
To begin, we set the initial state of the brain to be an activation pattern consistent with those empirically observed in the brain's baseline condition. More specifically, we set the initial state such that the regions of the default mode network had activity magnitudes equal to $1$ (``on''), while all other regions had activity magnitudes equal to $0$ (``off''). Furthermore, we examined 3 distinct target states such that regions of the (i) auditory, (ii) extended visual, or (iii) motor systems had activity magnitudes equal to $1$ (``on''), while all other regions had activity magnitudes equal to $0$ (``off''). In this context, we sought to understand characteristics of the transitions between initial and target states that could be performed with minimal energy, minimal time, and along short trajectories in state space by multiple control regions (multi-point control; see Fig.~\ref{fig:figure1}C and Methods). We note that mathematically, we measure time in arbitrary units, at each of which control energy can be utilized by a brain region. Intuitively, we operationalize time as consistent with the temporal scale at which brain regions can alter their activity magnitudes to affect state transitions.

\subsection*{Characteristics of Optimal Control Trajectories}

We first study the three state transitions from the default mode to (i) auditory, (ii) extended visual, and (iii) motor states (Fig.~2A). We take a hypothesis-driven approach and define the ``control set'' to be composed of dorsal and ventral attention \citep{posner1989attention}, fronto-parietal, and cingulo-opercular cognitive control regions \citep{gu2015controllability}. That is, this set of 87 regions will utilize control energy using a multi-point control strategy, thereby changing the time-varying activity magnitudes of all brain regions (Fig.~2B). The optimal trajectories display multiple peaks in the distance from the target state as a function of time, and are altered very little by whether the target state is the auditory, extended visual, or motor system (Fig.~2C). Because the optimal trajectory is determined via a balance of control energy and trajectory distance (see Methods), it stands to reason that the time-dependent energy utilized by the control set is inversely related to the distance between the current state and the target state. When little control energy is utilized, the current state can drift far from the target state, while when a larger magnitude of control energy is utilized, the current state moves closer to the target state (Fig.~2D).

It is important to note that these general characteristics of the optimal control trajectories are dependent on our choice of the control set (which here we guide with biologically motivated hypotheses), as well as on a penalty on the time required for the transition ($\rho$ in Equation[\ref{eqn_opt1}]; see Methods). In the supplement, we examine the effect of alternative choices for both the control set and $\rho$. First, we find that when the control set includes every node in the network, the distance to the target state decreases monotonically to zero along the trajectory (Fig.~S1A). Second, we consider the effect of the penalty term on control energy, $\rho$. For the results presented here, we fix $\rho$ to be equal to $1$. However, in the supplement, we explore a wide range of $\rho$ values, and show that when $\rho$ is small, the optimal control trajectory is largely driven by a minimization of the integrated squared distance to the target. In contrast, when $\rho$ is large, the optimal control trajectory is largely driven by the magnitude of the utilized energy (Fig.~S1B). Importantly, we did not perform a full sweep of $\rho$ from $0$ to infinity because very small values of $\rho$ cause numeric instabilities in the calculations.

\begin{figure}
\centering
\includegraphics[width=0.7\linewidth]{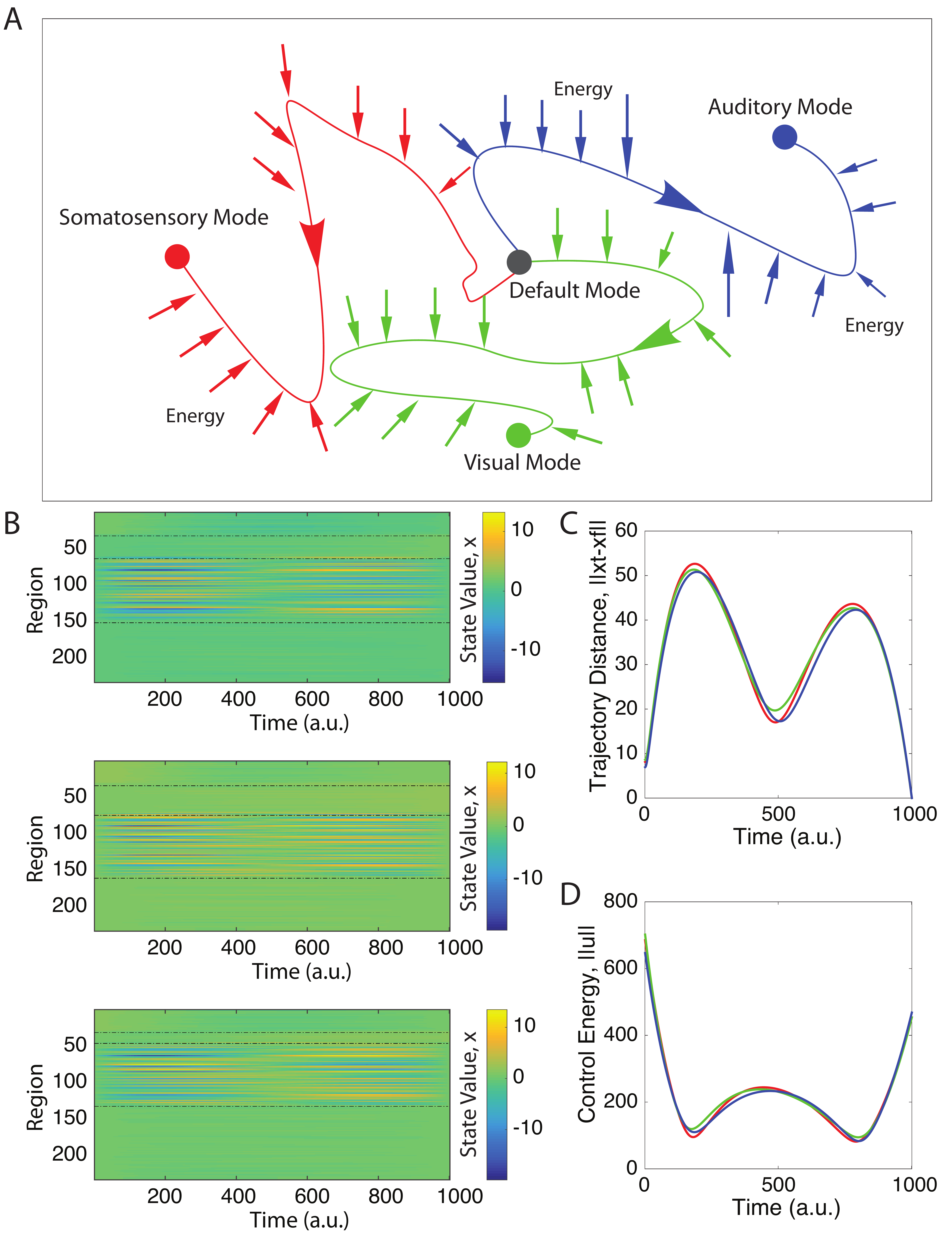}
\caption{{\bf Optimal Control Trajectories.} \emph{(A)} We study 3 distinct types of state transitions in which the initial state is characterized by high activity in the default mode system, and the target states are characterized by high activity in auditory (blue), extended visual (green), or motor (red) systems. \emph{(B)} The activation profiles of all $N=234$ brain regions as a function of time along the optimal control trajectory, illustrating that activity magnitudes vary by region and by time. Activation can be either positive or negative and the exact range of values will depend on the initial state, the target state, and the control set. Regions are listed in the following order: initial state, target state, controllers and others. \emph{(C)} The average distance from the current state $x(t)$ to the target state $x(T)$ as a function of time for the trajectories from the default mode system to the auditory, visual, and motor systems, illustrating behavior in the large state space. \emph{(D)} The average control energy utilized by the control set as a function of time for the trajectories from the default mode system to the auditory, visual, and motor systems. The similarity of the curves observed in panels \emph{(C)} and \emph{(D)} is driven largely by the fact that they share the same control set. See Fig. S2(B) for additional information on the range of these control energy values along the trajectories. Colors representing target states are identical in panels \emph{(A)}, \emph{(C)}, and \emph{(D)}.}
\label{fig:figure2}
\end{figure}


\subsection*{Structurally-Driven Task Preference for Control Regions}
We next ask whether certain brain regions are located at specific points in the structural network that make them predisposed to play consistent and important roles in driving optimal control trajectories. To answer this question, we choose control sets of the same size as the brain's hypothesized cognitive control set; recall that in the previous section, we defined the brain's cognitive control set to consist of the 87 nodes of the dorsal and ventral attention, fronto-parietal, and cingulo-opercular systems
following \citep{gu2015controllability}. Here, we choose the 87 regions of these new control sets uniformly at random from the set of all nodes. Using these ``random'' control sets, we computed the optimal control trajectory for each of the three state transitions and for each subject separately. Then, we rank the random control sets in descending order according to the energy cost of the trajectory and we assign every region participating in an $r$-ranked control set with rank-value $r$.  Next, we define the \emph{control efficiency} of a brain region to be the sum of its rank values in all of the random control sets it belongs divided by the total number of sets it belongs to. Intuitively, a region with a high control efficiency is one that exerts control with little energy utilization. Importantly, it must decrease activation in the initial state, and increase activation in the target state, a pair of capabilities that depend on the pattern of connections emanating from the region.

In general, we observe that a region's preference for being an optimal controller (exerting control with little energy utilization) is positively correlated with its network communicability to the regions of high activity in the target state (Spearman correlation $r = 0.27, p < 4.8 \times 10^{-4}$; see Fig.~3A). We recall that network communicability is a measurement of the strength of a connection from one region to another that accounts for walks of all lengths (see Methods). Interestingly, we observed this same correlation
between control efficiency and network communicability across optimal control trajectories for all three state transitions, from the default mode to the auditory ($r=0.36$, $p=1.4 \times 10^{-8}$), extended visual ($r=0.51$, $p=1.1 \times 10^{-16}$), or motor ($r=0.42$, $p=2.1 \times 10^{-11}$) systems (Fig.~3B-D). Together, these results indicate that regions that are close (in terms of walk lengths) to regions of high activity in the target state are efficient controllers for that specific state transition. Note that these regions are not purely target areas, likely due to the fact that they must also decrease activation in the initial state.

The general role that network proximity to the target state plays for control regions ensures that regions that are proximate to all three target states (auditory, extended visual, and motor) will be consistent controllers, while regions that are proximate to only one of the target states will be task-specific controllers. To better understand the anatomy of efficient controllers, we transformed control efficiency values to $z$-scores and defined an efficient control hub to be any region whose associated $p$-value was less than $0.025$. Across all three state transitions, we found that the supramarginal gyrus specifically, and the inferior parietal lobule more generally, consistently acted as efficient control hubs. The consistent control role of these regions is likely due to the fact that these areas are structurally interconnected with ventral premotor cortex, a key input to primary sensorimotor areas \citep{kandel2000principles}. The areas that are more specific to the three state transitions include medial parietal cortex (motor transition), orbitofrontal and inferior temporal cortex (visual transition), and superior temporal cortex (auditory transition).

\begin{figure}
\centering
\includegraphics[width=1.0\linewidth]{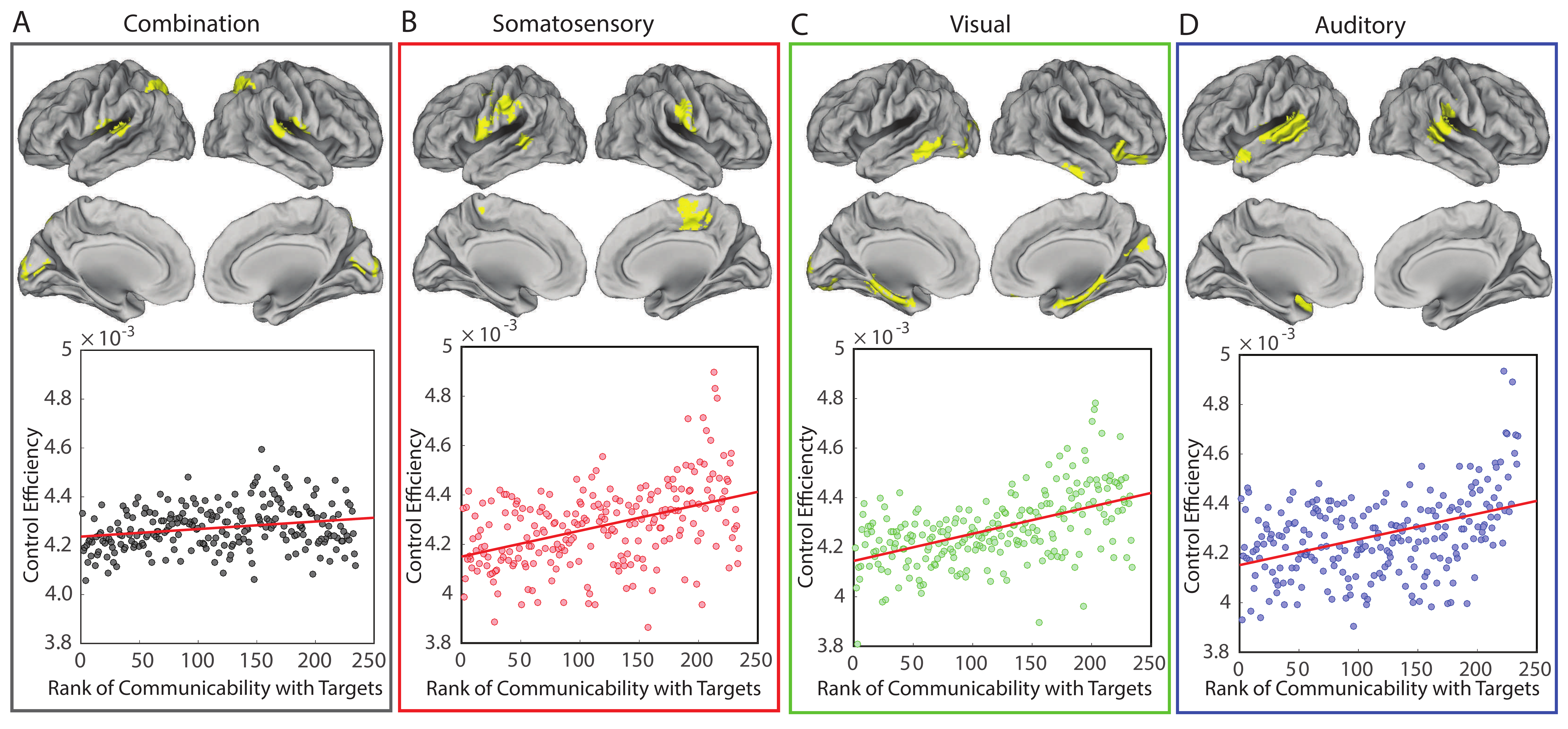}
\caption{ {\bf Structurally-Driven Task Preference for Control Regions.} \emph{(A)} \emph{Top} Regions with high control efficiency (see Eqn \ref{eqn:CPS}) across all 3 state transitions: from the default mode to auditory, extended visual, and motor systems. \emph{Bottom} Scatterplot of the control efficiency with the average network communicability to all 3 target regions (Spearman correlation $r = 0.27, p < 4.8 \times 10^{-4}$). \emph{(B--D)} \emph{Top} Regions with high control efficiency for the transition from default mode to \emph{(B)} motor, \emph{(C)} extended visual, and \emph{(D)} auditory ($r=0.36$, $p=1.4 \times 10^{-8}$) targets (\emph{top}). \emph{Bottom} Scatter plot of control efficiency \emph{versus} normalized network communicability with regions that are active in the target state: motor ($r=0.42$, $p=2.1 \times 10^{-11}$), extended visual ($r=0.51$, $p=1.1 \times 10^{-16}$), and auditory ($r=0.36$, $p=1.4 \times 10^{-8}$). Values of control efficiency in all four panels are averaged of subjects.}
\label{fig:figure3}
\end{figure}

\subsection*{Regional Roles in Control Tasks}

The analyses outlined above are built on the assumption that the brain uses fronto-parietal, cingulo-opercular, and attention systems to affect cognitive control, which we define as the ability to move the brain from an initial state (e.g., the default mode systems) to a specified final state (e.g., activation of extended visual, auditory, or motor cortex). However, one might naturally ask whether these regions of the brain could have been predicted \emph{a priori} to be effective controllers based on traditional engineering-based notions of control. In the control theory literature, particularly the literature devoted to the subfield of \emph{network controllability}, there exist several controllability notions, including average, modal, and boundary control \citep{pasqualetti2014controllability}. Average controllability identifies brain areas that can theoretically steer the system into many different states, or patterns of neurophysiological activity magnitudes across brain regions. Modal controllability identifies brain areas that can theoretically steer the system into difficult-to-reach states. Boundary controllability identifies brain areas that can theoretically steer the system into states where different cognitive systems are either coupled or decoupled. See the SI for mathematical definitions and \citep{gu2015controllability} for prior studies in human neuroimaging.

We calculated average, modal, and boundary control values for each node in the network. We observe that while cognitive control regions cover a broad swath of frontal and parietal cortex, including medial frontal cortex and anterior cingulate (Fig.~4A), the number of these regions that intersect with the strongest 87 average, modal, or boundary control hubs was on average approximately 50 (Fig.~4B).  These results suggest that the control capabilities of the human brain's cognitive control regions may not be perfectly aligned with control notions previously developed in the field of mechanical engineering, provided that the model assumptions and data quality are appropriate (see Methodological Considerations). Instead, cognitive control regions in the human brain may have distinct capabilities necessary for the specific transitions required by the brain under the constraints imposed by neuroanatomy and neurophysiology.

To more directly test this possibility, we examined the average distance (Fig.~4C) and energy
(Fig.~4D) for transitions from the default mode to the auditory, extended visual, and sensorimotor states that are driven by average, modal, and boundary control hubs, or by regions of fronto-parietal, cingulo-opercular, and attention systems. We observed that both the trajectory cost and the energy cost differ by control strategy and by target state. We quantify this observation using a 2-way ANOVA with both the control strategy and target state as categorical factors. Using the trajectory cost as the dependent variable, we observed a significant main effect of control strategy ($F = 78.74$, $p=4.65\times 10^{-41}$), a significant main effect of target state ($F=29.24$, $p =1.12 \times 10^{-12}$), and a significant interaction between control strategy and target state ($F=11.36$, $p=7.6 \times 10^{-12}$). Similarly, using the energy cost as the dependent variable, we observed a significant main effect of control strategy ($F =67.94$, $p=2.48 \times 10^{-36}$), a significant main effect of target state ($F=39.18$, $p =1.99 \times 10^{-16}$), and a significant interaction between control strategy and target state ($F=10.93$, $p=2.18 \times 10^{-11}$). Collapsing over target states and performing post-hoc testing, we observed that cognitive control regions displayed a similar average trajectory cost to average control hubs, but a lower average trajectory cost than modal and boundary control hubs ($p<0.05$ uncorrected). Furthermore, cognitive control regions possessed a higher average energy cost than the average and modal control hubs, but a lower average energy cost than the boundary control hubs. These results interestingly suggest that the human's cognitive control regions, as defined by decades of research in cognitive neuroscience, may affect state transitions using neither the shortest distances nor the lowest energies possible, provided that the model assumptions and data quality are appropriate (see Methodological Considerations). This is likely due to the fact that cognitive control regions must affect a broad array of state transitions that cannot easily be classified into average, modal, and boundary control strategies.

\begin{figure}
\centering
\includegraphics[width=1.0\linewidth]{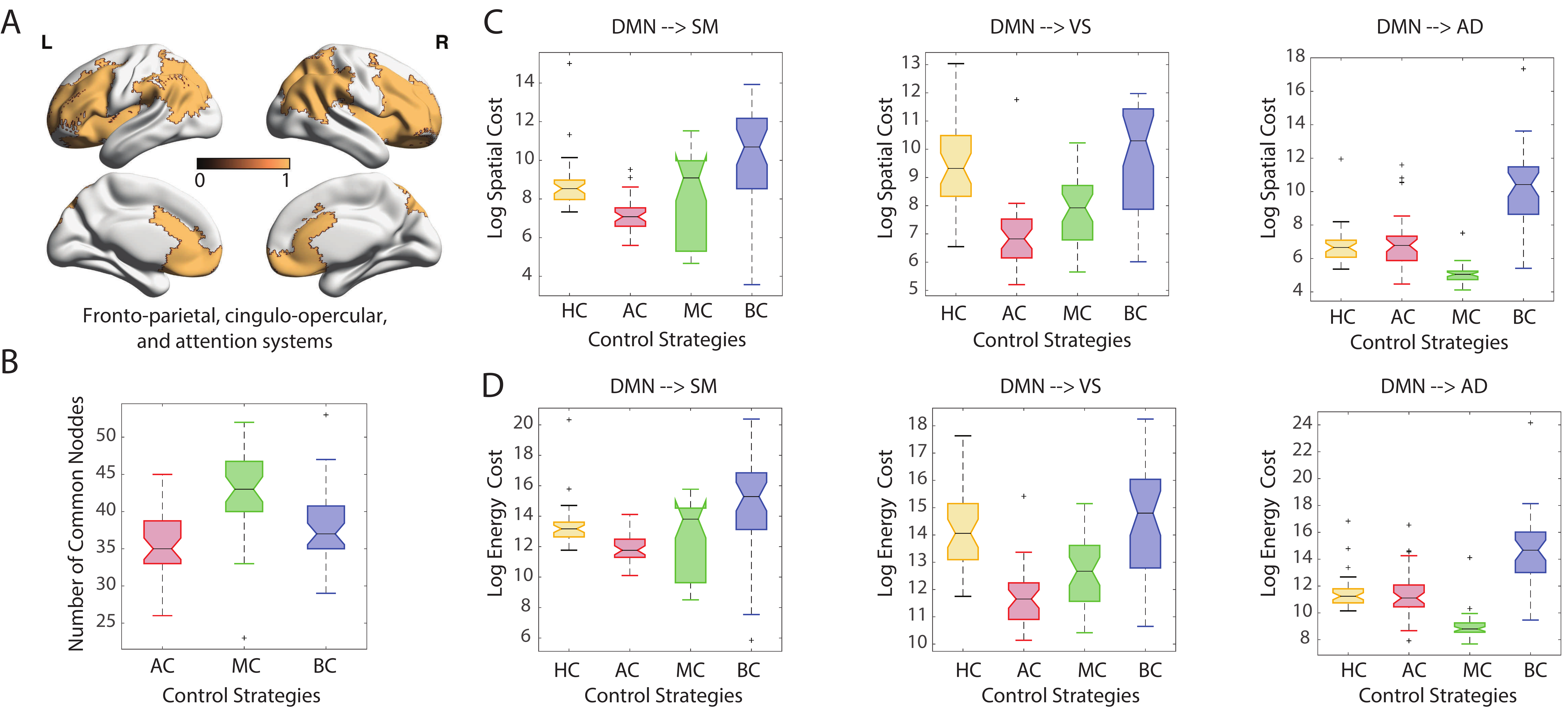}
\caption{{\bf Regional Roles in Control Tasks.} \emph{(A)} Cognitive control regions cover a broad swath of frontal and parietal cortex, including medial frontal cortex and anterior cingulate, and are defined as regions included in fronto-parietal, cingulo-opercular, and attention systems \citep{gu2015controllability}. \emph{(B)} The number of these regions overlapping with the strongest 87 average, modal and boundary control hubs is approximately 50. Different choices of control strategies result in variation in both \emph{(C)} trajectory cost and \emph{(D)} energy cost. Here, HC refers to cognitive control regions, AC refers to average control hubs, MC refers to modal control hubs, and BC refers to boundary control hubs. }
\label{fig:figure4}
\end{figure}


\subsection*{Specificity of Control in Health and Following Injury}

The unique role of brain regions in affecting control strategies may bring with it vulnerability to injury. When a brain network is injured, regional control roles may be significantly altered, potentially increasing susceptibility to underlying abnormalities in neuronal dynamics. To characterize this vulnerability, we determine the degree to which a single brain region impacts putative control processes and we ask whether that specificity is maintained or altered following brain injury. We measure specificity by iteratively removing nodes from the control set, and we compute the \emph{energetic impact} of each region on the optimal trajectory as the resulting increase in the log value of the energy cost (see Fig.~5A and Eqn \ref{eqn:rc} in Methods). Intuitively, regions with high energetic impact are those whose removal from the network causes the greatest increase in the energy required for the state transition. Across all subjects and all tasks, we observe that the regions with the highest energy impact are the supramarginal gyrus specifically, and the inferior parietal lobule more generally, the same regions that emerged as consistent and efficient controllers in Fig.~\ref{fig:figure2}A.

Next we determined whether energetic impact -- our proxy for regional specificity of control roles -- is altered in individuals with mild traumatic brain injury (mTBI). Intuitively, if all regions of a brain have high energetic impact, this indicates that each region is performing a different control role which is destroyed by removal of the node. By contrast, if all regions of a brain have low average energetic impact, this indicates that each region is performing a similar control role that is not destroyed by removal of a node. We observed that individuals with mTBI displayed anatomically similar patterns of energetic impact on control trajectories as regions are removed from the network (Fig.~5B). However, the average magnitude and variability of the energetic impact differed significantly between the two groups, with individuals having experienced mTBI displaying significantly lower values of average magnitude of energetic impact (permutation test: $p=5.0 \times 10^{-6}$) and lower values of the average standard deviation of energetic impact ($p=2.0 \times 10^{-6}$). We note that common graphic metrics including the degree, path length, clustering coefficient, modularity, local efficiency, global efficiency, and density were not significantly different between the two groups, suggesting that this effect is specific to control (see Supplement). These results indicate that mTBI patients display a loss of specificity in the putative control roles of brain regions, suggesting greater susceptibility to damage-induced noise in neurophysiological processes, or to external drivers in the form of stimulation.

\begin{figure}
\centering
\includegraphics[width=0.7\linewidth]{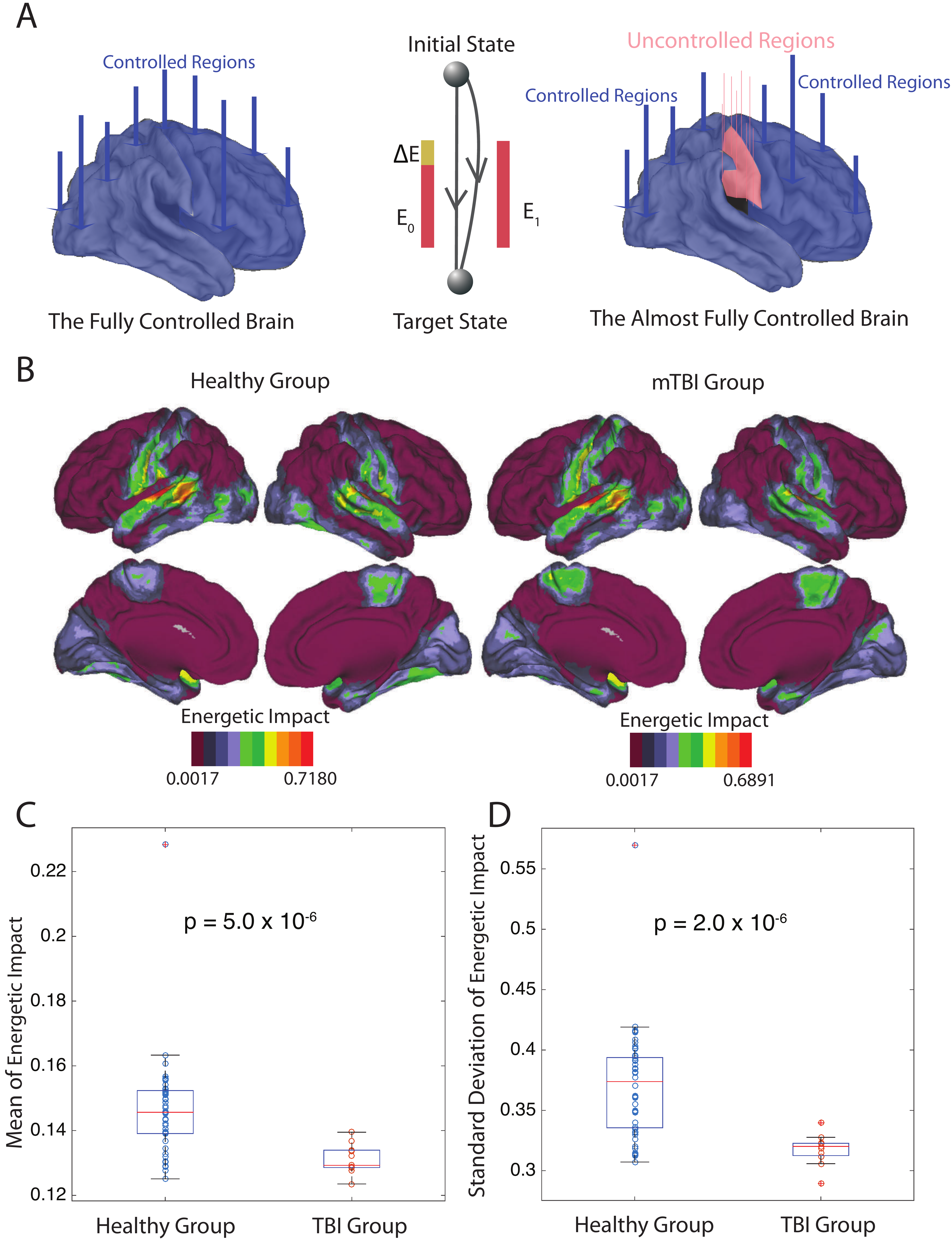}
\caption{{\bf Specificity of Control in Health and Following Injury.} \emph{(A)} Theoretically, the brain is fully controllable when every region is a control point, but may not be fully controllable when fewer regions are used to affect control. \emph{(B)} The regions with the highest values of energetic impact on control trajectories upon removal from the network, on average across subjects and tasks, were the supramarginal gyrus specifically, and the inferior parietal lobule more generally.  In general, the healthy group and the mTBI group displayed similar anatomical patterns of energetic impact.  \emph{(C)} Magnitude and standard derivation of energetic impact averaged over regions and tasks; boxplots indicate variation over subjects. Even after removing the single outlier in the healthy group, patients with mTBI displayed significantly lower values of average magnitude of energetic impact (permutation test: $p=1.1 \times 10^{-5}$) and lower values of the average standard deviation of energetic impact ($p=2.0 \times 10^{-6}$) than healthy controls.}
\label{fig:figure5}
\end{figure}

\section*{Discussion}
Here we ask whether structural connectivity forms a fundamental constraint on how the brain may move between diverse cognitive states. To address this question, we capitalize on recent advances in network control theory to identify and characterize optimal trajectories from an initial state (composed of high activity in the default mode system) to target states (composed of high activity in sensorimotor systems) with finite time, limited energy, and multi-point control. Using structural brain networks estimated from diffusion imaging data acquired in a large cohort of 48 healthy individuals and 11 patients with mild traumatic brain injury, we show that these optimal control trajectories are characterized by continuous changes in regional activity across the brain.  We show that the regions critical for eliciting these state transitions differ depending on the target state, but that heteromodal association hubs -- predominantly in the supramarginal gyrus specifically, and the inferior parietal lobule more generally -- are consistently recruited for all three transitions. Finally, we study the sensitivity of optimal control trajectories to the removal of nodes from the network, and we demonstrate that brain networks from individuals with mTBI display maladaptive control capabilities suggestive of a limited dynamic range of states available to the system. Together, these results offer initial insights into how structural network differences between individuals impact their potential to control transitions between cognitive states.

\subsection*{Role of Structural Connectivity in Shaping Brain Functional Patterns}

A growing body of literature on the relationship between brain structure and function has demonstrated that the brain's network of anatomical connections constrains the range of spontaneous \citep{deco2011emerging} and task-related \citep{hermundstad2013structural} fluctuations in brain activity. Evidence for such structural underpinnings comes from two distinct lines of research. On one hand, empirical studies have demonstrated that structural insults in the form of lesions result in acute reorganization of the brain's pattern of functional coupling \citep{johnston2008loss, o2013causal}. These observations are further buttressed by simulation studies in which structural connectivity has been used to constrain interactions among dynamic elements in biophysical models of brain activity
\citep{honey2007network, honey2009predicting, adachi2011functional} and models of network communication \citep{goni2014resting, abdelnour2014network, mivsic2015cooperative}. Though this forward modeling approach has proven fruitful in predicting observed patterns of functional connectivity, the precise mapping of brain structure to function remains unclear.

The present study builds on this body of work, using a dynamical model of how brain activity propagates over a network in order to gain insight into what features of that network facilitate easy transitions from a baseline (default mode) state to states where the brain's primary sensorimotor systems are activated. In contrast to previous simulation studies that have focused on network features that influence the passive spread of activity over time, this present study directly engages the question of how those same features enable the state of the system to be controlled. We use this model to demonstrate that brain regions are differentially-suited for particular control tasks, roles that can be predicted on the basis of how well-connected they are to regions in the target state. Regions that are close (in terms of walk lengths) to regions of high activity in the target state are efficient controllers for that specific state transition. It follows, then, that a brain region's capacity to dynamically influence a network depends not only on its pattern of connectivity, but also the repertoire of states that the system visits. In other words, a region that maintains many connections (both direct and indirect), but never to regions that are ``active'' in target states, may exert less influence than a region that maintains few connections, but whose connections are distributed among regions that are ``active'' in many target states. We further demonstrate that this mapping of brain structure to specific functions is altered in individuals with mTBI, suggesting that injury may alter control profiles of individual brain regions.

Our finding that the inferior parietal lobule forms a consistently effective control region, across all three target states, is particularly interesting when considered in the context of prior literature on this region's structural and functional roles. In particular, the inferior parietal lobule represents the superior portion of the temporoparietal junction, a multimodal area associated with functions as wide ranging as calculation, finger gnosis, left/right orientation, and writing \citep{rusconi2009disconnection}. Focal damage to this area leads to \rev{wide-spread cognitive disfunction} (e.g., Gerstmann syndrome) as a consequence of the unique confluence of white matter pathways underlying this region \citep{rushworth2006connection}. The diverse white matter projections emanating from this area may support its putative role in effectively controlling brain function. Indeed, recent evidence suggests that the right temporoparietal junction links two antagonistic brain networks processing external \emph{versus} internal information: a midcingulate-motor-insular network associated with attention, and a parietal network associated with social cognition and memory retrieval \citep{bzdok2013characterization}. These data support the notion that the right temporoparietal junction controls our attention to salient external events \citep{corbetta2002control}, perhaps with early input from the right fronto-insular cortex thought to drive switching between central-executive and default-mode networks \citep{sridharan2008critical}.

\subsection*{Single \emph{versus} Multipoint Control}

An important feature of our model lies in the delineation of a control \emph{set}, a group of brain regions that can affect distributed control. The focus on multiple points of control throughout the system is one that has important theoretical motivations and empirical correlates. Prior computational models demonstrate that while the brain is theoretically controllable via input to a single control point, the energy and time required for that control is such that the brain is practically uncontrollable \citep{gu2015controllability}. These data argue for an assessment of multi-point control as a better proxy of control strategies that the brain might utilize. Indeed, such an argument is consistent with empirical observations that stimulation (or even drug manipulations) focused on single brain regions are less effective in treating psychiatric disease than interventions that target multiple brain regions \citep{Sommer2012,Tortella2014}. A prime empirical example of multi-point control is cognitive behavioral therapy, which offers a spatio-temporal pattern of activations that enhances cognitive function and decreases psychiatric symptoms across diagnostic categories \citep{Lett2014,Radhu2012,Cima2014}. Other potential multi-point control mechanisms include grid stimulation across multiple electrodes, suggested in the control of medically refractory epilepsy \citep{Ching2012}.

\subsection*{Diversity in Human Brain Control Strategies}

By studying multi-point control, we were able to directly assess whether the optimal control trajectories elicited from fronto-parietal, cingulo-opercular, and attention systems displayed similar distance and energy to trajectories to those obtained using regions selected for engineering-based notions of control \citep{pasqualetti2014controllability}. Specifically, we compare and contrast the performance of human cognitive control regions to average, modal, and boundary controllers \cite{gu2015controllability}. One might naturally ask whether and how each of these engineering-based notions of controllers is an appropriate theoretical quantity to consider in the context of neurobiology.  Average controllers are those theoretically able to push the system from any arbitrary initial state to any easily-reachable state, nearby on the energy landscape. Modal controllers are those that are optimally placed to move the system from any arbitrary initial state to any difficult-to-reach state, far away on the energy landscape. Boundary controllers are those that are optimally placed to integrate or segregate network communities in the system. Common assumptions that underlie each of these control strategies are that (i) controllers can be identified independently from the initial and final states, and (ii) all states in the energy landscape are accessible to the system. The common constraint on each of these strategies is that expended energy must be minimized for a region to be referred to as a \emph{controller}. In the human brain, it is not well-known whether these assumptions are met, or whether the energetic constraint is sufficient to predict control functions.

Interestingly, we observe that the optimal trajectories elicited by canonically defined cognitive control regions do not show similar energy requirements or trajectory distances to any of these previously described control types. There are several potential reasons for this observation: (i) false negatives in the diffusion imaging data impacting on the observed network profiles, (ii) assumptions of the linear model, and (iii) \emph{bona fide} differences between mechanical controllers and biological controllers. Prior work demonstrating robustness of controllability profiles across large cohorts and different diffusion imaging acquisition protocols provide initial evidence that the first explanation is unlikely to fully explain our findings \citep{gu2015controllability}. Regarding the second explanation -- assumptions of the linear model -- it is interesting to note that recent evidence suggests that the average, modal, and boundary controllability profiles identified by the linear model provide excellent predictions for the behavior of nonlinear models \citep{muldoon2016stimulation}. Evidence supporting the third interpretation -- that these observed differences are \emph{bona fide} differences between mechanical controllers and biological controllers -- is provided by the fact that cognitive control regions must affect a broad array of state transitions that do not easily fit into prior classifications. These transitions include switching behavior \citep{Hansen2015}, inter-state competition \citep{cocchi2013dynamic}, distributed rather than centralized control \citep{eisenreich2016control}, and push-pull control \citep{Khambhati2016}, which may each offer differential advantages for neural computations \citep{Durstewitz2008}.

\subsection*{Maladaptive Control in Traumatic Brain Injury}

Finally, our assessment of patients with mild traumatic brain injury enabled us to determine whether widespread injury leads to a decrement in the healthy network control profiles, thus theoretically requiring greater energy for the same functions, or an enhancement of the healthy network control profiles at the cost of expected sensitivity to external perturbations. Our data provide initial evidence for maladaptive control of the latter sort in patients with mild traumatic brain injury. Understanding the impact of brain injury on cognitive processes, including the ability to switch between cognitive states, is a major goal in clinical neuroscience. Indeed, traumatic brain injury is a common source of brain dysfunction, affecting more than 200,000 individuals per year in the United States alone.  Injuries -- often caused by motor vehicle and sports accidents -- result in damage to neuronal axons, including long-distance white matter fiber bundles \citep{Johnson2013} as well as u-fibers and deep white matter tracks with multiple crossings. The pattern of injury can be multi-focal and variable across individuals
\citep{Kinnunen2011,Sidaros2008,Hellyer2013}, challenging comprehensive predictors and generalizable interventions.

Recent evidence suggests that injury-induced, widespread damage to white matter tracts critically impacts large-scale network organization in the human brain, as measured by diffusion imaging tractography \citep{Kinnunen2011,Fagerholm2015}. Moreover, this damage is associated with fundamental changes in cognitive function \citep{Sharp2014}, including information processing speed, executive function, and associative memory \citep{Fagerholm2015}. Each of these cognitive deficits intuitively depends on the ability to transition from one cognitive state to another; yet an understanding of structural drivers of these transitions and their potential alteration in mTBI has remained elusive. Here we demonstrate a loss of specificity in putative control processes in mTBI, suggesting that the unique roles of individual brain regions in supporting cognitive state transitions are damaged. It is intuitively plausible that this decrement in regional specificity of control leads to broad changes in functional dynamics, particularly in the system's susceptibility to damage-induced noise in neurophysiological processes \citep{Garrett2013}. Indeed, the observed decrements in energetic impact might further provide a direct structural mechanism for the decreased signal variability observed in mTBI using electrophysiological imaging \citep{Raja2012,Nenadovic2008}. More generally, these findings highlight the fact that the healthy brain might display a degree of controllability that is either decremented or enhanced in injury and disease, suggesting the possibility of a U-shaped curve reminiscent of similar curves observed in other brain network phenotypes \citep{collin2013ontegeny,cools2011inverted}.

\subsection*{Methodological Considerations}

A few methodological points are worthy of additional consideration.  First, in this study we examined structural brain networks derived from diffusion imaging data and associated tractography algorithms. These algorithms remain in their relative infancy, and can still report spurious tracts or fail to report existing tracts \citep{thomas2014anatomical, reveley2015superficial,pestilli2014evaluation}. Despite the evolving nature of diffusion protocols and tractography algorithms, preliminary data provide initial evidence that consistent controllability profiles can be robustly observed across large cohorts and different diffusion imaging acquisition protocols \citep{gu2015controllability}. Formal validation in axonal tracing studies in monkeys and other mammals \citep{jbabdi2013topographic} remains the gold standard for these types of data. \rev{However, it is important to note that initial work supports the notion that much of the structure present in DSI connectivity matrices recapitulates known projections observed in tract tracing studies of the macaque \cite{hagmann2008mapping}.}

Second, following \citep{gu2015controllability,betzel2016optimally,muldoon2016stimulation}, we employ a linear dynamical model, consistent with prior empirical studies demonstrating their ability to predict features of resting state fMRI data \citep{galan2008network,honey2009predicting}.  This choice is to some degree predicated on the well-developed theoretical and analytical results in the engineering and physics literatures examining the relationship between control and network topology \citep{liu2011controllability, muller2011few, yan2012controlling}. Moreover, it is plausible that even these results using simple linear models may offer important intuitions for controlling nonlinear models of brain function. Indeed, theoretical work over the last several decades has demonstrated the utility of describing non-linear systems in terms
of a linear approximation in the neighborhood of the system's equilibrium points \citep{luenberger1979introduction}. Very recent evidence has extended these intuitions to neuroimaging data, demonstrating that the average, modal, and boundary controllability profiles identified by the linear model can be used to predict the behavior of nonlinear models in the form of Wilson-Cowan oscillators, which are commonly used to understand the dynamics of cortical columns \citep{muldoon2016stimulation}.

\subsection*{Future directions}

An interesting hypothesis generated by the current framework is that control capabilities may be altered dimensionally across traditionally separated diagnostic groups that display dysconnectivity in network hubs, as measured by regions of high eigenvector centrality.  Such a hypothesis builds on the now seminal \emph{dysconnection hypothesis} in schizophrenia \citep{stephan2009dysconnection}, and expands it to include an explicit dynamical control component. Indeed, mounting evidence suggests that the overload or failure of brain network hubs may be a common neurophysiological mechanism of a
range of neurological disorders including Alzheimer's disease, multiple sclerosis, traumatic brain injury and epilepsy \citep{Stam2014}. Alterations in these hubs can also be used to predict the progression of psychiatric disorders such as schizophrenia \citep{Collin2015}. In both neurological and psychiatric disorders, these changes to network hubs may alter the control capabilities of the individual, challenging the normal executive functions required for daily living. It is also intuitively plausible that normal variation in hub architecture may play a role in individual differences in control capabilities in healthy individuals, impacting on the speed with which they transition between cognitive states. These topics will form important provender for future work.

\section*{Acknowledgements:}
D.S.B., S.G., and R.F.B. acknowledge support from the John D. and Catherine T. MacArthur Foundation, the Alfred P. Sloan Foundation, the Army Research Laboratory and the Army Research Office through contract numbers W911NF-10-2-0022 and W911NF-14-1-0679,the National Institute of Mental Health (2-R01-DC-009209-11), the National Institute of Child Health and Human Development (1R01HD086888-01), the Office of Naval Research, and the National Science Foundation (BCS-1441502, BCS-1430087, and CAREER PHY-1554488). S.T.G. and P.R.D. acknowledge support from a Head Health Challenge grant from General Electric and the National Football League. F.P. acknowledges support from the National Science Foundation award \#BCS 1430280. The content is solely the responsibility of the authors and does not necessarily represent the official views of any of the funding agencies.
\section*{Author Contributions} S.G. performed the analysis. M.C. preprocessed the data. S.T.G. and P.R.D. acquired the data. S.G., R.F.B., F.P., D.S.B. developed the project. S.G., R.F.B., F.P., D.S.B. wrote the paper.\\
\section*{Competing Interests} The authors declare that they have no competing financial interests.\\
\section*{References}

\bibliography{dynamical_trajectories,dynamical_trajectories_v5_discussionOnly_db,dynamical_trajectories_v5_structureFunction_RB,bibfile,bibfile_2,bibfile_new,bibfile_original}

\begin{thebibliography}{102}
\expandafter\ifx\csname natexlab\endcsname\relax\def\natexlab#1{#1}\fi
\providecommand{\url}[1]{\texttt{#1}}
\providecommand{\href}[2]{#2}
\providecommand{\path}[1]{#1}
\providecommand{\DOIprefix}{doi:}
\providecommand{\ArXivprefix}{arXiv:}
\providecommand{\URLprefix}{URL: }
\providecommand{\Pubmedprefix}{pmid:}
\providecommand{\doi}[1]{\href{http://dx.doi.org/#1}{\path{#1}}}
\providecommand{\Pubmed}[1]{\href{pmid:#1}{\path{#1}}}
\providecommand{\bibinfo}[2]{#2}
\ifx\xfnm\relax \def\xfnm[#1]{\unskip,\space#1}\fi
\bibitem[{Abdelnour et~al.(2014)Abdelnour, Voss \& Raj}]{abdelnour2014network}
\bibinfo{author}{Abdelnour, F.}, \bibinfo{author}{Voss, H.~U.}, \&
  \bibinfo{author}{Raj, A.} (\bibinfo{year}{2014}).
\newblock \bibinfo{title}{Network diffusion accurately models the relationship
  between structural and functional brain connectivity networks}.
\newblock {\it \bibinfo{journal}{Neuroimage}\/},  {\it \bibinfo{volume}{90}\/},
  \bibinfo{pages}{335--347}.
\bibitem[{Adachi et~al.(2011)Adachi, Osada, Sporns, Watanabe, Matsui, Miyamoto
  \& Miyashita}]{adachi2011functional}
\bibinfo{author}{Adachi, Y.}, \bibinfo{author}{Osada, T.},
  \bibinfo{author}{Sporns, O.}, \bibinfo{author}{Watanabe, T.},
  \bibinfo{author}{Matsui, T.}, \bibinfo{author}{Miyamoto, K.}, \&
  \bibinfo{author}{Miyashita, Y.} (\bibinfo{year}{2011}).
\newblock \bibinfo{title}{Functional connectivity between anatomically
  unconnected areas is shaped by collective network-level effects in the
  macaque cortex}.
\newblock {\it \bibinfo{journal}{Cerebral cortex}\/},  (p.
  \bibinfo{pages}{bhr234}).
\bibitem[{Alavash et~al.(2015)Alavash, Hilgetag, Thiel \&
  Gie{\ss}ing}]{alavash2015persistency}
\bibinfo{author}{Alavash, M.}, \bibinfo{author}{Hilgetag, C.~C.},
  \bibinfo{author}{Thiel, C.~M.}, \& \bibinfo{author}{Gie{\ss}ing, C.}
  (\bibinfo{year}{2015}).
\newblock \bibinfo{title}{Persistency and flexibility of complex brain networks
  underlie dual-task interference}.
\newblock {\it \bibinfo{journal}{Human brain mapping}\/},  {\it
  \bibinfo{volume}{36}\/}, \bibinfo{pages}{3542--3562}.
\bibitem[{Attwell \& Laughlin(2001)}]{Attwell2001}
\bibinfo{author}{Attwell, D.}, \& \bibinfo{author}{Laughlin, S.~B.}
  (\bibinfo{year}{2001}).
\newblock \bibinfo{title}{An energy budget for signaling in the grey matter of
  the brain}.
\newblock {\it \bibinfo{journal}{J Cereb Blood Flow Metab}\/},  {\it
  \bibinfo{volume}{21}\/}, \bibinfo{pages}{1133--1145}.
\bibitem[{Bassett et~al.(2011{\natexlab{a}})Bassett, Brown, Deshpande, Carlson
  \& Grafton}]{bassett2011conserved}
\bibinfo{author}{Bassett, D.~S.}, \bibinfo{author}{Brown, J.~A.},
  \bibinfo{author}{Deshpande, V.}, \bibinfo{author}{Carlson, J.~M.}, \&
  \bibinfo{author}{Grafton, S.~T.} (\bibinfo{year}{2011}{\natexlab{a}}).
\newblock \bibinfo{title}{Conserved and variable architecture of human white
  matter connectivity}.
\newblock {\it \bibinfo{journal}{Neuroimage}\/},  {\it \bibinfo{volume}{54}\/},
  \bibinfo{pages}{1262--1279}.
\bibitem[{Bassett et~al.(2010)Bassett, Greenfield, Meyer-Lindenberg,
  Weinberger, Moore \& Bullmore}]{bassett2010efficient}
\bibinfo{author}{Bassett, D.~S.}, \bibinfo{author}{Greenfield, D.~L.},
  \bibinfo{author}{Meyer-Lindenberg, A.}, \bibinfo{author}{Weinberger, D.~R.},
  \bibinfo{author}{Moore, S.~W.}, \& \bibinfo{author}{Bullmore, E.~T.}
  (\bibinfo{year}{2010}).
\newblock \bibinfo{title}{Efficient physical embedding of topologically complex
  information processing networks in brains and computer circuits}.
\newblock {\it \bibinfo{journal}{PLoS Comput Biol}\/},  {\it
  \bibinfo{volume}{6}\/}, \bibinfo{pages}{e1000748}.
\bibitem[{Bassett et~al.(2011{\natexlab{b}})Bassett, Wymbs, Porter, Mucha,
  Carlson \& Grafton}]{Bassett2011}
\bibinfo{author}{Bassett, D.~S.}, \bibinfo{author}{Wymbs, N.~F.},
  \bibinfo{author}{Porter, M.~A.}, \bibinfo{author}{Mucha, P.~J.},
  \bibinfo{author}{Carlson, J.~M.}, \& \bibinfo{author}{Grafton, S.~T.}
  (\bibinfo{year}{2011}{\natexlab{b}}).
\newblock \bibinfo{title}{Dynamic reconfiguration of human brain networks
  during learning}.
\newblock {\it \bibinfo{journal}{Proc Natl Acad Sci U S A}\/},  {\it
  \bibinfo{volume}{108}\/}, \bibinfo{pages}{7641--7646}.
\bibitem[{Bassett et~al.(2013)Bassett, Wymbs, Rombach, Porter, Mucha \&
  Grafton}]{Bassett2013}
\bibinfo{author}{Bassett, D.~S.}, \bibinfo{author}{Wymbs, N.~F.},
  \bibinfo{author}{Rombach, M.~P.}, \bibinfo{author}{Porter, M.~A.},
  \bibinfo{author}{Mucha, P.~J.}, \& \bibinfo{author}{Grafton, S.~T.}
  (\bibinfo{year}{2013}).
\newblock \bibinfo{title}{Task-based core-periphery organization of human brain
  dynamics}.
\newblock {\it \bibinfo{journal}{PLoS Comput Biol}\/},  {\it
  \bibinfo{volume}{9}\/}, \bibinfo{pages}{e1003171}.
\bibitem[{Bassett et~al.(2015)Bassett, Yang, Wymbs \& Grafton}]{Bassett2015}
\bibinfo{author}{Bassett, D.~S.}, \bibinfo{author}{Yang, M.},
  \bibinfo{author}{Wymbs, N.~F.}, \& \bibinfo{author}{Grafton, S.~T.}
  (\bibinfo{year}{2015}).
\newblock \bibinfo{title}{Learning-induced autonomy of sensorimotor systems}.
\newblock {\it \bibinfo{journal}{Nat Neurosci}\/},  {\it
  \bibinfo{volume}{18}\/}, \bibinfo{pages}{744--751}.
\bibitem[{Betzel et~al.(2016)Betzel, Gu, Medaglia, Pasqualetti \&
  Bassett}]{betzel2016optimally}
\bibinfo{author}{Betzel, R.~F.}, \bibinfo{author}{Gu, S.},
  \bibinfo{author}{Medaglia, J.~D.}, \bibinfo{author}{Pasqualetti, F.}, \&
  \bibinfo{author}{Bassett, D.~S.} (\bibinfo{year}{2016}).
\newblock \bibinfo{title}{Optimally controlling the human connectome: the role
  of network topology}.
\newblock {\it \bibinfo{journal}{arXiv preprint arXiv:1603.05261}\/}, .
\bibitem[{Boltyanskii et~al.(1960)Boltyanskii, Gamkrelidze \&
  Pontryagin}]{boltyanskii1960theory}
\bibinfo{author}{Boltyanskii, V.~G.}, \bibinfo{author}{Gamkrelidze, R.~V.}, \&
  \bibinfo{author}{Pontryagin, L.~S.} (\bibinfo{year}{1960}).
\newblock {\it \bibinfo{title}{The theory of optimal processes. I. The maximum
  principle}\/}.
\newblock \bibinfo{type}{Technical Report} DTIC Document.
\bibitem[{Braun et~al.(2015)Braun, Muldoon \& Bassett}]{braun2015human}
\bibinfo{author}{Braun, U.}, \bibinfo{author}{Muldoon, S.~F.}, \&
  \bibinfo{author}{Bassett, D.~S.} (\bibinfo{year}{2015}).
\newblock \bibinfo{title}{On human brain networks in health and disease}.
\newblock {\it \bibinfo{journal}{eLS}\/}, .
\bibitem[{Bullmore \& Sporns(2009)}]{bullmore2009complex}
\bibinfo{author}{Bullmore, E.}, \& \bibinfo{author}{Sporns, O.}
  (\bibinfo{year}{2009}).
\newblock \bibinfo{title}{Complex brain networks: graph theoretical analysis of
  structural and functional systems}.
\newblock {\it \bibinfo{journal}{Nature Reviews Neuroscience}\/},  {\it
  \bibinfo{volume}{10}\/}, \bibinfo{pages}{186--198}.
\bibitem[{Bzdok et~al.(2013)Bzdok, Langner, Schilbach, Jakobs, Roski, Caspers,
  Laird, Fox, Zilles \& Eickhoff}]{bzdok2013characterization}
\bibinfo{author}{Bzdok, D.}, \bibinfo{author}{Langner, R.},
  \bibinfo{author}{Schilbach, L.}, \bibinfo{author}{Jakobs, O.},
  \bibinfo{author}{Roski, C.}, \bibinfo{author}{Caspers, S.},
  \bibinfo{author}{Laird, A.~R.}, \bibinfo{author}{Fox, P.~T.},
  \bibinfo{author}{Zilles, K.}, \& \bibinfo{author}{Eickhoff, S.~B.}
  (\bibinfo{year}{2013}).
\newblock \bibinfo{title}{Characterization of the temporo-parietal junction by
  combining data-driven parcellation, complementary connectivity analyses, and
  functional decoding}.
\newblock {\it \bibinfo{journal}{Neuroimage}\/},  {\it \bibinfo{volume}{81}\/},
  \bibinfo{pages}{381--392}.
\bibitem[{Cammoun et~al.(2012)Cammoun, Gigandet, Meskaldji, Thiran, Sporns, Do,
  Maeder, Meuli \& Hagmann}]{Cammoun2012}
\bibinfo{author}{Cammoun, L.}, \bibinfo{author}{Gigandet, X.},
  \bibinfo{author}{Meskaldji, D.}, \bibinfo{author}{Thiran, J.~P.},
  \bibinfo{author}{Sporns, O.}, \bibinfo{author}{Do, K.~Q.},
  \bibinfo{author}{Maeder, P.}, \bibinfo{author}{Meuli, R.}, \&
  \bibinfo{author}{Hagmann, P.} (\bibinfo{year}{2012}).
\newblock \bibinfo{title}{Mapping the human connectome at multiple scales with
  diffusion spectrum {MRI}}.
\newblock {\it \bibinfo{journal}{J Neurosci Methods}\/},  {\it
  \bibinfo{volume}{203}\/}, \bibinfo{pages}{386--397}.
\bibitem[{Carter et~al.(2012)Carter, Patel, Astafiev, Snyder, Rengachary,
  Strube, Pope, Shimony, Lang, Shulman \& Corbetta}]{Carter2012}
\bibinfo{author}{Carter, A.~R.}, \bibinfo{author}{Patel, K.~R.},
  \bibinfo{author}{Astafiev, S.~V.}, \bibinfo{author}{Snyder, A.~Z.},
  \bibinfo{author}{Rengachary, J.}, \bibinfo{author}{Strube, M.~J.},
  \bibinfo{author}{Pope, A.}, \bibinfo{author}{Shimony, J.~S.},
  \bibinfo{author}{Lang, C.~E.}, \bibinfo{author}{Shulman, G.~L.}, \&
  \bibinfo{author}{Corbetta, M.} (\bibinfo{year}{2012}).
\newblock \bibinfo{title}{Upstream dysfunction of somatomotor functional
  connectivity after corticospinal damage in stroke}.
\newblock {\it \bibinfo{journal}{Neurorehabil Neural Repair}\/},  {\it
  \bibinfo{volume}{26}\/}, \bibinfo{pages}{7--19}.
\bibitem[{Ching et~al.(2012)Ching, Brown \& Kramer}]{Ching2012}
\bibinfo{author}{Ching, S.}, \bibinfo{author}{Brown, E.~N.}, \&
  \bibinfo{author}{Kramer, M.~A.} (\bibinfo{year}{2012}).
\newblock \bibinfo{title}{Distributed control in a mean-field cortical network
  model: implications for seizure suppression}.
\newblock {\it \bibinfo{journal}{Phys Rev E Stat Nonlin Soft Matter Phys}\/},
  {\it \bibinfo{volume}{86}\/}, \bibinfo{pages}{021920}.
\bibitem[{Cieslak \& Grafton(2014)}]{cieslak2014local}
\bibinfo{author}{Cieslak, M.}, \& \bibinfo{author}{Grafton, S.}
  (\bibinfo{year}{2014}).
\newblock \bibinfo{title}{Local termination pattern analysis: a tool for
  comparing white matter morphology}.
\newblock {\it \bibinfo{journal}{Brain imaging and behavior}\/},  {\it
  \bibinfo{volume}{8}\/}, \bibinfo{pages}{292--299}.
\bibitem[{Cima et~al.(2014)Cima, Andersson, Schmidt \& Henry}]{Cima2014}
\bibinfo{author}{Cima, R.~F.}, \bibinfo{author}{Andersson, G.},
  \bibinfo{author}{Schmidt, C.~J.}, \& \bibinfo{author}{Henry, J.~A.}
  (\bibinfo{year}{2014}).
\newblock \bibinfo{title}{Cognitive-behavioral treatments for tinnitus: a
  review of the literature}.
\newblock {\it \bibinfo{journal}{J Am Acad Audiol}\/},  {\it
  \bibinfo{volume}{25}\/}, \bibinfo{pages}{29--61}.
\bibitem[{Cocchi et~al.(2013)Cocchi, Zalesky, Fornito \&
  Mattingley}]{cocchi2013dynamic}
\bibinfo{author}{Cocchi, L.}, \bibinfo{author}{Zalesky, A.},
  \bibinfo{author}{Fornito, A.}, \& \bibinfo{author}{Mattingley, J.~B.}
  (\bibinfo{year}{2013}).
\newblock \bibinfo{title}{Dynamic cooperation and competition between brain
  systems during cognitive control}.
\newblock {\it \bibinfo{journal}{Trends in cognitive sciences}\/},  {\it
  \bibinfo{volume}{17}\/}, \bibinfo{pages}{493--501}.
\bibitem[{Collin \& van~den Heuvel(2013)}]{collin2013ontegeny}
\bibinfo{author}{Collin, G.}, \& \bibinfo{author}{van~den Heuvel, M.~P.}
  (\bibinfo{year}{2013}).
\newblock \bibinfo{title}{The ontogeny of the human connectome: development and
  dynamic changes of brain connectivity across the life span}.
\newblock {\it \bibinfo{journal}{Neuroscientist}\/},  {\it
  \bibinfo{volume}{19}\/}, \bibinfo{pages}{616--628}.
\bibitem[{Collin et~al.(2015)Collin, de~Nijs, Hulshoff~Pol, Cahn \& van~den
  Heuvel}]{Collin2015}
\bibinfo{author}{Collin, G.}, \bibinfo{author}{de~Nijs, J.},
  \bibinfo{author}{Hulshoff~Pol, H.~E.}, \bibinfo{author}{Cahn, W.}, \&
  \bibinfo{author}{van~den Heuvel, M.~P.} (\bibinfo{year}{2015}).
\newblock \bibinfo{title}{Connectome organization is related to longitudinal
  changes in general functioning, symptoms and {IQ} in chronic schizophrenia}.
\newblock {\it \bibinfo{journal}{Schizophr Res}\/},  {\it
  \bibinfo{volume}{S0920--9964}\/}, \bibinfo{pages}{00141--00143}.
\bibitem[{Cools \& D'Esposito(2011)}]{cools2011inverted}
\bibinfo{author}{Cools, R.}, \& \bibinfo{author}{D'Esposito, M.}
  (\bibinfo{year}{2011}).
\newblock \bibinfo{title}{Inverted-{U}-shaped dopamine actions on human working
  memory and cognitive control}.
\newblock {\it \bibinfo{journal}{Biol Psychiatry}\/},  {\it
  \bibinfo{volume}{69}\/}, \bibinfo{pages}{e113--e125}.
\bibitem[{Corbetta \& Shulman(2002)}]{corbetta2002control}
\bibinfo{author}{Corbetta, M.}, \& \bibinfo{author}{Shulman, G.~L.}
  (\bibinfo{year}{2002}).
\newblock \bibinfo{title}{Control of goal-directed and stimulus-driven
  attention in the brain}.
\newblock {\it \bibinfo{journal}{Nat Rev Neurosci}\/},  {\it
  \bibinfo{volume}{3}\/}, \bibinfo{pages}{201--215}.
\bibitem[{Crofts \& Higham(2009)}]{crofts2009weighted}
\bibinfo{author}{Crofts, J.~J.}, \& \bibinfo{author}{Higham, D.~J.}
  (\bibinfo{year}{2009}).
\newblock \bibinfo{title}{A weighted communicability measure applied to complex
  brain networks}.
\newblock {\it \bibinfo{journal}{Journal of the Royal Society Interface}\/},
  (pp. \bibinfo{pages}{rsif--2008}).
\bibitem[{Daducci et~al.(2012)Daducci, Gerhard, Griffa, Lemkaddem, Cammoun,
  Gigandet, Meuli, Hagmann \& Thiran}]{Daducci2012}
\bibinfo{author}{Daducci, A.}, \bibinfo{author}{Gerhard, S.},
  \bibinfo{author}{Griffa, A.}, \bibinfo{author}{Lemkaddem, A.},
  \bibinfo{author}{Cammoun, L.}, \bibinfo{author}{Gigandet, X.},
  \bibinfo{author}{Meuli, R.}, \bibinfo{author}{Hagmann, P.}, \&
  \bibinfo{author}{Thiran, J.~P.} (\bibinfo{year}{2012}).
\newblock \bibinfo{title}{The connectome mapper: an open-source processing
  pipeline to map connectomes with {MRI}}.
\newblock {\it \bibinfo{journal}{PLoS One}\/},  {\it \bibinfo{volume}{7}\/},
  \bibinfo{pages}{e48121}.
\bibitem[{Deco et~al.(2011)Deco, Jirsa \& McIntosh}]{deco2011emerging}
\bibinfo{author}{Deco, G.}, \bibinfo{author}{Jirsa, V.~K.}, \&
  \bibinfo{author}{McIntosh, A.~R.} (\bibinfo{year}{2011}).
\newblock \bibinfo{title}{Emerging concepts for the dynamical organization of
  resting-state activity in the brain}.
\newblock {\it \bibinfo{journal}{Nat Rev Neurosci}\/},  {\it
  \bibinfo{volume}{12}\/}, \bibinfo{pages}{43--56}.
\bibitem[{Di~Martino et~al.(2014)Di~Martino, Fair, Kelly, Satterthwaite,
  Castellanos, Thomason, Craddock, Luna, Leventhal, Zuo
  et~al.}]{di2014unraveling}
\bibinfo{author}{Di~Martino, A.}, \bibinfo{author}{Fair, D.~A.},
  \bibinfo{author}{Kelly, C.}, \bibinfo{author}{Satterthwaite, T.~D.},
  \bibinfo{author}{Castellanos, F.~X.}, \bibinfo{author}{Thomason, M.~E.},
  \bibinfo{author}{Craddock, R.~C.}, \bibinfo{author}{Luna, B.},
  \bibinfo{author}{Leventhal, B.~L.}, \bibinfo{author}{Zuo, X.-N.} et~al.
  (\bibinfo{year}{2014}).
\newblock \bibinfo{title}{Unraveling the miswired connectome: a developmental
  perspective}.
\newblock {\it \bibinfo{journal}{Neuron}\/},  {\it \bibinfo{volume}{83}\/},
  \bibinfo{pages}{1335--1353}.
\bibitem[{Durstewitz \& Deco(2008)}]{Durstewitz2008}
\bibinfo{author}{Durstewitz, D.}, \& \bibinfo{author}{Deco, G.}
  (\bibinfo{year}{2008}).
\newblock \bibinfo{title}{Computational significance of transient dynamics in
  cortical networks}.
\newblock {\it \bibinfo{journal}{Eur J Neurosci}\/},  {\it
  \bibinfo{volume}{27}\/}, \bibinfo{pages}{217--227}.
\bibitem[{Eisenreich et~al.(2016)Eisenreich, Akaishi \&
  Hayden}]{eisenreich2016control}
\bibinfo{author}{Eisenreich, B.}, \bibinfo{author}{Akaishi, R.}, \&
  \bibinfo{author}{Hayden, B.} (\bibinfo{year}{2016}).
\newblock \bibinfo{title}{Control without controllers: Towards a distributed
  neuroscience of executive control}.
\newblock {\it \bibinfo{journal}{bioRxiv}\/},  {\it \bibinfo{volume}{1101}\/},
  \bibinfo{pages}{077685}.
\bibitem[{Fagerholm et~al.(2015)Fagerholm, Hellyer, Scott, Leech \&
  Sharp}]{Fagerholm2015}
\bibinfo{author}{Fagerholm, E.~D.}, \bibinfo{author}{Hellyer, P.~J.},
  \bibinfo{author}{Scott, G.}, \bibinfo{author}{Leech, R.}, \&
  \bibinfo{author}{Sharp, D.~J.} (\bibinfo{year}{2015}).
\newblock \bibinfo{title}{Disconnection of network hubs and cognitive
  impairment after traumatic brain injury}.
\newblock {\it \bibinfo{journal}{Brain}\/},  {\it \bibinfo{volume}{138}\/},
  \bibinfo{pages}{1696--1709}.
\bibitem[{Fiete et~al.(2010)Fiete, Senn, Wang \& Hahnloser}]{Fiete2010}
\bibinfo{author}{Fiete, I.~R.}, \bibinfo{author}{Senn, W.},
  \bibinfo{author}{Wang, C.~Z.}, \& \bibinfo{author}{Hahnloser, R. H.~R.}
  (\bibinfo{year}{2010}).
\newblock \bibinfo{title}{Spike-time-dependent plasticity and heterosynaptic
  competition organize networks to produce long scale-free sequences of neural
  activity}.
\newblock {\it \bibinfo{journal}{Neuron}\/},  {\it \bibinfo{volume}{65}\/},
  \bibinfo{pages}{563--576}.
\bibitem[{Freeman(1994)}]{freeman1994characterization}
\bibinfo{author}{Freeman, W.~J.} (\bibinfo{year}{1994}).
\newblock \bibinfo{title}{Characterization of state transitions in spatially
  distributed, chaotic, nonlinear, dynamical systems in cerebral cortex}.
\newblock {\it \bibinfo{journal}{Integr Physiol Behav Sci}\/},  {\it
  \bibinfo{volume}{29}\/}, \bibinfo{pages}{294--306}.
\bibitem[{Gal{\'a}n(2008)}]{galan2008network}
\bibinfo{author}{Gal{\'a}n, R.~F.} (\bibinfo{year}{2008}).
\newblock \bibinfo{title}{On how network architecture determines the dominant
  patterns of spontaneous neural activity}.
\newblock {\it \bibinfo{journal}{PLoS One}\/},  {\it \bibinfo{volume}{3}\/},
  \bibinfo{pages}{e2148}.
\bibitem[{Garrett et~al.(2013)Garrett, Samanez-Larkin, MacDonald, Lindenberger,
  McIntosh \& Grady}]{Garrett2013}
\bibinfo{author}{Garrett, D.~D.}, \bibinfo{author}{Samanez-Larkin, G.~R.},
  \bibinfo{author}{MacDonald, S.~W.}, \bibinfo{author}{Lindenberger, U.},
  \bibinfo{author}{McIntosh, A.~R.}, \& \bibinfo{author}{Grady, C.~L.}
  (\bibinfo{year}{2013}).
\newblock \bibinfo{title}{Moment-to-moment brain signal variability: a next
  frontier in human brain mapping?}
\newblock {\it \bibinfo{journal}{Neurosci Biobehav Rev}\/},  {\it
  \bibinfo{volume}{37}\/}, \bibinfo{pages}{610--624}.
\bibitem[{Gazzaniga(2013)}]{Gazzaniga2013}
\bibinfo{editor}{Gazzaniga, M.~S.} (Ed.) (\bibinfo{year}{2013}).
\newblock {\it \bibinfo{title}{The cognitive neurosciences}\/}.
\newblock \bibinfo{publisher}{MIT Press}.
\bibitem[{Go{\~n}i et~al.(2014)Go{\~n}i, van~den Heuvel, Avena-Koenigsberger,
  de~Mendizabal, Betzel, Griffa, Hagmann, Corominas-Murtra, Thiran \&
  Sporns}]{goni2014resting}
\bibinfo{author}{Go{\~n}i, J.}, \bibinfo{author}{van~den Heuvel, M.~P.},
  \bibinfo{author}{Avena-Koenigsberger, A.}, \bibinfo{author}{de~Mendizabal,
  N.~V.}, \bibinfo{author}{Betzel, R.~F.}, \bibinfo{author}{Griffa, A.},
  \bibinfo{author}{Hagmann, P.}, \bibinfo{author}{Corominas-Murtra, B.},
  \bibinfo{author}{Thiran, J.-P.}, \& \bibinfo{author}{Sporns, O.}
  (\bibinfo{year}{2014}).
\newblock \bibinfo{title}{Resting-brain functional connectivity predicted by
  analytic measures of network communication}.
\newblock {\it \bibinfo{journal}{Proceedings of the National Academy of
  Sciences}\/},  {\it \bibinfo{volume}{111}\/}, \bibinfo{pages}{833--838}.
\bibitem[{Gu et~al.(2016)Gu, Cieslak, Baird, Muldoon, Grafton, Pasqualetti \&
  Bassett}]{gu2016energy}
\bibinfo{author}{Gu, S.}, \bibinfo{author}{Cieslak, M.},
  \bibinfo{author}{Baird, B.}, \bibinfo{author}{Muldoon, S.~F.},
  \bibinfo{author}{Grafton, S.~T.}, \bibinfo{author}{Pasqualetti, F.}, \&
  \bibinfo{author}{Bassett, D.~S.} (\bibinfo{year}{2016}).
\newblock \bibinfo{title}{The energy landscape of neurophysiological activity
  implicit in brain network structure}.
\newblock {\it \bibinfo{journal}{Submitted}\/}, .
\bibitem[{Gu et~al.(2015)Gu, Pasqualetti, Cieslak, Telesford, Alfred, Kahn,
  Medaglia, Vettel, Miller, Grafton et~al.}]{gu2015controllability}
\bibinfo{author}{Gu, S.}, \bibinfo{author}{Pasqualetti, F.},
  \bibinfo{author}{Cieslak, M.}, \bibinfo{author}{Telesford, Q.~K.},
  \bibinfo{author}{Alfred, B.~Y.}, \bibinfo{author}{Kahn, A.~E.},
  \bibinfo{author}{Medaglia, J.~D.}, \bibinfo{author}{Vettel, J.~M.},
  \bibinfo{author}{Miller, M.~B.}, \bibinfo{author}{Grafton, S.~T.} et~al.
  (\bibinfo{year}{2015}).
\newblock \bibinfo{title}{Controllability of structural brain networks}.
\newblock {\it \bibinfo{journal}{Nature communications}\/},  {\it
  \bibinfo{volume}{6}\/}.
\bibitem[{Hagmann et~al.(2008)Hagmann, Cammoun, Gigandet, Meuli, Honey, Wedeen
  \& Sporns}]{hagmann2008mapping}
\bibinfo{author}{Hagmann, P.}, \bibinfo{author}{Cammoun, L.},
  \bibinfo{author}{Gigandet, X.}, \bibinfo{author}{Meuli, R.},
  \bibinfo{author}{Honey, C.~J.}, \bibinfo{author}{Wedeen, V.~J.}, \&
  \bibinfo{author}{Sporns, O.} (\bibinfo{year}{2008}).
\newblock \bibinfo{title}{Mapping the structural core of human cerebral
  cortex}.
\newblock {\it \bibinfo{journal}{PLoS Biol}\/},  {\it \bibinfo{volume}{6}\/},
  \bibinfo{pages}{e159}.
\bibitem[{Hansen et~al.(2015)Hansen, Battaglia, Spiegler, Deco \&
  Jirsa}]{Hansen2015}
\bibinfo{author}{Hansen, E.~C.}, \bibinfo{author}{Battaglia, D.},
  \bibinfo{author}{Spiegler, A.}, \bibinfo{author}{Deco, G.}, \&
  \bibinfo{author}{Jirsa, V.~K.} (\bibinfo{year}{2015}).
\newblock \bibinfo{title}{Functional connectivity dynamics: modeling the
  switching behavior of the resting state}.
\newblock {\it \bibinfo{journal}{Neuroimage}\/},  {\it
  \bibinfo{volume}{105}\/}, \bibinfo{pages}{525--535}.
\bibitem[{Hellyer et~al.(2013)Hellyer, Leech, Ham, Bonnelle \&
  Sharp}]{Hellyer2013}
\bibinfo{author}{Hellyer, P.~J.}, \bibinfo{author}{Leech, R.},
  \bibinfo{author}{Ham, T.~E.}, \bibinfo{author}{Bonnelle, V.}, \&
  \bibinfo{author}{Sharp, D.~J.} (\bibinfo{year}{2013}).
\newblock \bibinfo{title}{Individual prediction of white matter injury
  following traumatic brain injury}.
\newblock {\it \bibinfo{journal}{Ann Neurol}\/},  {\it \bibinfo{volume}{73}\/},
  \bibinfo{pages}{489--499}.
\bibitem[{Hermundstad et~al.(2013{\natexlab{a}})Hermundstad, Bassett, Brown,
  Aminoff, Clewett, Freeman, Frithsen, Johnson, Tipper, Miller
  et~al.}]{Hermundstad2013}
\bibinfo{author}{Hermundstad, A.~M.}, \bibinfo{author}{Bassett, D.~S.},
  \bibinfo{author}{Brown, K.~S.}, \bibinfo{author}{Aminoff, E.~M.},
  \bibinfo{author}{Clewett, D.}, \bibinfo{author}{Freeman, S.},
  \bibinfo{author}{Frithsen, A.}, \bibinfo{author}{Johnson, A.},
  \bibinfo{author}{Tipper, C.~M.}, \bibinfo{author}{Miller, M.~B.} et~al.
  (\bibinfo{year}{2013}{\natexlab{a}}).
\newblock \bibinfo{title}{Structural foundations of resting-state and
  task-based functional connectivity in the human brain}.
\newblock {\it \bibinfo{journal}{Proceedings of the National Academy of
  Sciences}\/},  {\it \bibinfo{volume}{110}\/}, \bibinfo{pages}{6169--6174}.
\bibitem[{Hermundstad et~al.(2013{\natexlab{b}})Hermundstad, Bassett, Brown,
  Aminoff, Clewett, Freeman, Frithsen, Johnson, Tipper, Miller
  et~al.}]{hermundstad2013structural}
\bibinfo{author}{Hermundstad, A.~M.}, \bibinfo{author}{Bassett, D.~S.},
  \bibinfo{author}{Brown, K.~S.}, \bibinfo{author}{Aminoff, E.~M.},
  \bibinfo{author}{Clewett, D.}, \bibinfo{author}{Freeman, S.},
  \bibinfo{author}{Frithsen, A.}, \bibinfo{author}{Johnson, A.},
  \bibinfo{author}{Tipper, C.~M.}, \bibinfo{author}{Miller, M.~B.} et~al.
  (\bibinfo{year}{2013}{\natexlab{b}}).
\newblock \bibinfo{title}{Structural foundations of resting-state and
  task-based functional connectivity in the human brain}.
\newblock {\it \bibinfo{journal}{Proceedings of the National Academy of
  Sciences}\/},  {\it \bibinfo{volume}{110}\/}, \bibinfo{pages}{6169--6174}.
\bibitem[{Hermundstad et~al.(2014{\natexlab{a}})Hermundstad, Brown, Bassett,
  Aminoff, Frithsen, Johnson, Tipper, Miller, Grafton \&
  Carlson}]{hermundstad2014structurally}
\bibinfo{author}{Hermundstad, A.~M.}, \bibinfo{author}{Brown, K.~S.},
  \bibinfo{author}{Bassett, D.~S.}, \bibinfo{author}{Aminoff, E.~M.},
  \bibinfo{author}{Frithsen, A.}, \bibinfo{author}{Johnson, A.},
  \bibinfo{author}{Tipper, C.~M.}, \bibinfo{author}{Miller, M.~B.},
  \bibinfo{author}{Grafton, S.~T.}, \& \bibinfo{author}{Carlson, J.~M.}
  (\bibinfo{year}{2014}{\natexlab{a}}).
\newblock \bibinfo{title}{Structurally-constrained relationships between
  cognitive states in the human brain}.
\newblock {\it \bibinfo{journal}{PLoS Comput Biol}\/},  {\it
  \bibinfo{volume}{10}\/}, \bibinfo{pages}{e1003591}.
\bibitem[{Hermundstad et~al.(2014{\natexlab{b}})Hermundstad, Brown, Bassett,
  Aminoff, Frithsen, Johnson, Tipper, Miller, Grafton \&
  Carlson}]{Hermundstad2014}
\bibinfo{author}{Hermundstad, A.~M.}, \bibinfo{author}{Brown, K.~S.},
  \bibinfo{author}{Bassett, D.~S.}, \bibinfo{author}{Aminoff, E.~M.},
  \bibinfo{author}{Frithsen, A.}, \bibinfo{author}{Johnson, A.},
  \bibinfo{author}{Tipper, C.~M.}, \bibinfo{author}{Miller, M.~B.},
  \bibinfo{author}{Grafton, S.~T.}, \& \bibinfo{author}{Carlson, J.~M.}
  (\bibinfo{year}{2014}{\natexlab{b}}).
\newblock \bibinfo{title}{Structurally-constrained relationships between
  cognitive states in the human brain}.
\newblock {\it \bibinfo{journal}{PLoS Comput Biol}\/},  {\it
  \bibinfo{volume}{10}\/}, \bibinfo{pages}{e1003591}.
\bibitem[{Hermundstad et~al.(2011)Hermundstad, Brown, Bassett \&
  Carlson}]{Hermundstad2011}
\bibinfo{author}{Hermundstad, A.~M.}, \bibinfo{author}{Brown, K.~S.},
  \bibinfo{author}{Bassett, D.~S.}, \& \bibinfo{author}{Carlson, J.~M.}
  (\bibinfo{year}{2011}).
\newblock \bibinfo{title}{Learning, memory, and the role of neural network
  architecture}.
\newblock {\it \bibinfo{journal}{PLoS Comput Biol}\/},  {\it
  \bibinfo{volume}{7}\/}, \bibinfo{pages}{e1002063}.
\bibitem[{Honey et~al.(2009)Honey, Sporns, Cammoun, Gigandet, Thiran, Meuli \&
  Hagmann}]{honey2009predicting}
\bibinfo{author}{Honey, C.}, \bibinfo{author}{Sporns, O.},
  \bibinfo{author}{Cammoun, L.}, \bibinfo{author}{Gigandet, X.},
  \bibinfo{author}{Thiran, J.-P.}, \bibinfo{author}{Meuli, R.}, \&
  \bibinfo{author}{Hagmann, P.} (\bibinfo{year}{2009}).
\newblock \bibinfo{title}{Predicting human resting-state functional
  connectivity from structural connectivity}.
\newblock {\it \bibinfo{journal}{Proceedings of the National Academy of
  Sciences}\/},  {\it \bibinfo{volume}{106}\/}, \bibinfo{pages}{2035--2040}.
\bibitem[{Honey et~al.(2007)Honey, K{\"o}tter, Breakspear \&
  Sporns}]{honey2007network}
\bibinfo{author}{Honey, C.~J.}, \bibinfo{author}{K{\"o}tter, R.},
  \bibinfo{author}{Breakspear, M.}, \& \bibinfo{author}{Sporns, O.}
  (\bibinfo{year}{2007}).
\newblock \bibinfo{title}{Network structure of cerebral cortex shapes
  functional connectivity on multiple time scales}.
\newblock {\it \bibinfo{journal}{Proceedings of the National Academy of
  Sciences}\/},  {\it \bibinfo{volume}{104}\/}, \bibinfo{pages}{10240--10245}.
\bibitem[{Jbabdi et~al.(2013)Jbabdi, Sotiropoulos \&
  Behrens}]{jbabdi2013topographic}
\bibinfo{author}{Jbabdi, S.}, \bibinfo{author}{Sotiropoulos, S.~N.}, \&
  \bibinfo{author}{Behrens, T.~E.} (\bibinfo{year}{2013}).
\newblock \bibinfo{title}{The topographic connectome}.
\newblock {\it \bibinfo{journal}{Current opinion in neurobiology}\/},  {\it
  \bibinfo{volume}{23}\/}, \bibinfo{pages}{207--215}.
\bibitem[{Johnson et~al.(2013)Johnson, Stewart \& Smith}]{Johnson2013}
\bibinfo{author}{Johnson, V.~E.}, \bibinfo{author}{Stewart, W.}, \&
  \bibinfo{author}{Smith, D.~H.} (\bibinfo{year}{2013}).
\newblock \bibinfo{title}{Axonal pathology in traumatic brain injury}.
\newblock {\it \bibinfo{journal}{Exp Neurol}\/},  {\it
  \bibinfo{volume}{246}\/}, \bibinfo{pages}{35--43}.
\bibitem[{Johnston et~al.(2008)Johnston, Vaishnavi, Smyth, Zhang, He, Zempel,
  Shimony, Snyder \& Raichle}]{johnston2008loss}
\bibinfo{author}{Johnston, J.~M.}, \bibinfo{author}{Vaishnavi, S.~N.},
  \bibinfo{author}{Smyth, M.~D.}, \bibinfo{author}{Zhang, D.},
  \bibinfo{author}{He, B.~J.}, \bibinfo{author}{Zempel, J.~M.},
  \bibinfo{author}{Shimony, J.~S.}, \bibinfo{author}{Snyder, A.~Z.}, \&
  \bibinfo{author}{Raichle, M.~E.} (\bibinfo{year}{2008}).
\newblock \bibinfo{title}{Loss of resting interhemispheric functional
  connectivity after complete section of the corpus callosum}.
\newblock {\it \bibinfo{journal}{The Journal of neuroscience}\/},  {\it
  \bibinfo{volume}{28}\/}, \bibinfo{pages}{6453--6458}.
\bibitem[{Kalpinski et~al.(2013)Kalpinski, Williamson, Elliott, Berry,
  Underhill \& Fine}]{Kalpinski2013}
\bibinfo{author}{Kalpinski, R.~J.}, \bibinfo{author}{Williamson, M.~L.},
  \bibinfo{author}{Elliott, T.~R.}, \bibinfo{author}{Berry, J.~W.},
  \bibinfo{author}{Underhill, A.~T.}, \& \bibinfo{author}{Fine, P.~R.}
  (\bibinfo{year}{2013}).
\newblock \bibinfo{title}{Modeling the prospective relationships of impairment,
  injury severity, and participation to quality of life following traumatic
  brain injury}.
\newblock {\it \bibinfo{journal}{Biomed Res Int}\/},  {\it
  \bibinfo{volume}{2013}\/}, \bibinfo{pages}{102570}.
\bibitem[{Kandel et~al.(2000)Kandel, Schwartz, Jessell
  et~al.}]{kandel2000principles}
\bibinfo{author}{Kandel, E.~R.}, \bibinfo{author}{Schwartz, J.~H.},
  \bibinfo{author}{Jessell, T.~M.} et~al. (\bibinfo{year}{2000}).
\newblock {\it \bibinfo{title}{Principles of neural science}\/}
  volume~\bibinfo{volume}{4}.
\newblock \bibinfo{publisher}{McGraw-hill New York}.
\bibitem[{Khambhati et~al.(2016)Khambhati, Davis, Lucas, Litt \&
  Bassett}]{Khambhati2016}
\bibinfo{author}{Khambhati, A.}, \bibinfo{author}{Davis, K.},
  \bibinfo{author}{Lucas, T.}, \bibinfo{author}{Litt, B.}, \&
  \bibinfo{author}{Bassett, D.~S.} (\bibinfo{year}{2016}).
\newblock \bibinfo{title}{Virtual cortical resection reveals push-pull network
  control preceding seizure evolution}.
\newblock {\it \bibinfo{journal}{Submitted}\/}, .
\bibitem[{Kinnunen et~al.(2011)Kinnunen, Greenwood, Powell, Leech, Hawkins,
  Bonnelle, Patel, Counsell \& Sharp}]{Kinnunen2011}
\bibinfo{author}{Kinnunen, K.~M.}, \bibinfo{author}{Greenwood, R.},
  \bibinfo{author}{Powell, J.~H.}, \bibinfo{author}{Leech, R.},
  \bibinfo{author}{Hawkins, P.~C.}, \bibinfo{author}{Bonnelle, V.},
  \bibinfo{author}{Patel, M.~C.}, \bibinfo{author}{Counsell, S.~J.}, \&
  \bibinfo{author}{Sharp, D.~J.} (\bibinfo{year}{2011}).
\newblock \bibinfo{title}{White matter damage and cognitive impairment after
  traumatic brain injury}.
\newblock {\it \bibinfo{journal}{Brain}\/},  {\it \bibinfo{volume}{134}\/},
  \bibinfo{pages}{449--463}.
\bibitem[{Klimm et~al.(2014)Klimm, Bassett, Carlson \&
  Mucha}]{klimm2014resolving}
\bibinfo{author}{Klimm, F.}, \bibinfo{author}{Bassett, D.~S.},
  \bibinfo{author}{Carlson, J.~M.}, \& \bibinfo{author}{Mucha, P.~J.}
  (\bibinfo{year}{2014}).
\newblock \bibinfo{title}{Resolving structural variability in network models
  and the brain}.
\newblock {\it \bibinfo{journal}{PLOS Comput Biol}\/},  {\it
  \bibinfo{volume}{10}\/}, \bibinfo{pages}{e1003491}.
\bibitem[{Laughlin(2001)}]{Laughlin2001}
\bibinfo{author}{Laughlin, S.~B.} (\bibinfo{year}{2001}).
\newblock \bibinfo{title}{Efficiency and complexity in neural coding}.
\newblock {\it \bibinfo{journal}{Novartis Found Symp}\/},  {\it
  \bibinfo{volume}{239}\/}, \bibinfo{pages}{177--187}.
\bibitem[{Laughlin et~al.(1998)Laughlin, de~Ruyter~van Steveninck \&
  Anderson}]{Laughlin1998}
\bibinfo{author}{Laughlin, S.~B.}, \bibinfo{author}{de~Ruyter~van Steveninck,
  R.~R.}, \& \bibinfo{author}{Anderson, J.~C.} (\bibinfo{year}{1998}).
\newblock \bibinfo{title}{The metabolic cost of neural information}.
\newblock {\it \bibinfo{journal}{Nat Neurosci}\/},  {\it
  \bibinfo{volume}{1}\/}, \bibinfo{pages}{36--41}.
\bibitem[{Lee et~al.(2011)Lee, Ueno \& Yamashita}]{Lee2011}
\bibinfo{author}{Lee, S.}, \bibinfo{author}{Ueno, M.}, \&
  \bibinfo{author}{Yamashita, T.} (\bibinfo{year}{2011}).
\newblock \bibinfo{title}{Axonal remodeling for motor recovery after traumatic
  brain injury requires downregulation of gamma-aminobutyric acid signaling}.
\newblock {\it \bibinfo{journal}{Cell Death Dis}\/},  {\it
  \bibinfo{volume}{2}\/}, \bibinfo{pages}{e133}.
\bibitem[{Lett et~al.(2014)Lett, Voineskos, Kennedy, Levine \&
  Daskalakis}]{Lett2014}
\bibinfo{author}{Lett, T.~A.}, \bibinfo{author}{Voineskos, A.~N.},
  \bibinfo{author}{Kennedy, J.~L.}, \bibinfo{author}{Levine, B.}, \&
  \bibinfo{author}{Daskalakis, Z.~J.} (\bibinfo{year}{2014}).
\newblock \bibinfo{title}{Treating working memory deficits in schizophrenia: a
  review of the neurobiology}.
\newblock {\it \bibinfo{journal}{Biol Psychiatry}\/},  {\it
  \bibinfo{volume}{75}\/}, \bibinfo{pages}{361--370}.
\bibitem[{Levy et~al.(2001)Levy, Horn, Meilijson \& Ruppin}]{Levy2001}
\bibinfo{author}{Levy, N.}, \bibinfo{author}{Horn, D.},
  \bibinfo{author}{Meilijson, I.}, \& \bibinfo{author}{Ruppin, E.}
  (\bibinfo{year}{2001}).
\newblock \bibinfo{title}{Distributed synchrony in a cell assembly of spiking
  neurons}.
\newblock {\it \bibinfo{journal}{Neural Netw}\/},  {\it
  \bibinfo{volume}{14}\/}, \bibinfo{pages}{815--824}.
\bibitem[{Liu et~al.(2011)Liu, Slotine \&
  Barab{\'a}si}]{liu2011controllability}
\bibinfo{author}{Liu, Y.-Y.}, \bibinfo{author}{Slotine, J.-J.}, \&
  \bibinfo{author}{Barab{\'a}si, A.-L.} (\bibinfo{year}{2011}).
\newblock \bibinfo{title}{Controllability of complex networks}.
\newblock {\it \bibinfo{journal}{Nature}\/},  {\it \bibinfo{volume}{473}\/},
  \bibinfo{pages}{167--173}.
\bibitem[{Luenberger(1979)}]{luenberger1979introduction}
\bibinfo{author}{Luenberger, D.} (\bibinfo{year}{1979}).
\newblock \bibinfo{title}{Introduction to dynamic systems: theory, models, and
  applications}, .
\bibitem[{Medaglia et~al.(2015)Medaglia, Lynall \& Bassett}]{Medaglia2015}
\bibinfo{author}{Medaglia, J.~D.}, \bibinfo{author}{Lynall, M.~E.}, \&
  \bibinfo{author}{Bassett, D.~S.} (\bibinfo{year}{2015}).
\newblock \bibinfo{title}{Cognitive network neuroscience}.
\newblock {\it \bibinfo{journal}{J Cogn Neurosci}\/},  {\it
  \bibinfo{volume}{27}\/}, \bibinfo{pages}{1471--1491}.
\bibitem[{Meunier et~al.(2009)Meunier, Achard, Morcom \&
  Bullmore}]{Meunier2009}
\bibinfo{author}{Meunier, D.}, \bibinfo{author}{Achard, S.},
  \bibinfo{author}{Morcom, A.}, \& \bibinfo{author}{Bullmore, E.}
  (\bibinfo{year}{2009}).
\newblock \bibinfo{title}{Age-related changes in modular organization of human
  brain functional networks}.
\newblock {\it \bibinfo{journal}{Neuroimage}\/},  {\it \bibinfo{volume}{44}\/},
  \bibinfo{pages}{715--723}.
\bibitem[{Mi{\v{s}}i{\'c} et~al.(2015)Mi{\v{s}}i{\'c}, Betzel, Nematzadeh,
  Go{\~n}i, Griffa, Hagmann, Flammini, Ahn \& Sporns}]{mivsic2015cooperative}
\bibinfo{author}{Mi{\v{s}}i{\'c}, B.}, \bibinfo{author}{Betzel, R.~F.},
  \bibinfo{author}{Nematzadeh, A.}, \bibinfo{author}{Go{\~n}i, J.},
  \bibinfo{author}{Griffa, A.}, \bibinfo{author}{Hagmann, P.},
  \bibinfo{author}{Flammini, A.}, \bibinfo{author}{Ahn, Y.-Y.}, \&
  \bibinfo{author}{Sporns, O.} (\bibinfo{year}{2015}).
\newblock \bibinfo{title}{Cooperative and competitive spreading dynamics on the
  human connectome}.
\newblock {\it \bibinfo{journal}{Neuron}\/},  {\it \bibinfo{volume}{86}\/},
  \bibinfo{pages}{1518--1529}.
\bibitem[{Muldoon et~al.(2016{\natexlab{a}})Muldoon, Bridgeford \&
  Bassett}]{muldoon2016small}
\bibinfo{author}{Muldoon, S.~F.}, \bibinfo{author}{Bridgeford, E.~W.}, \&
  \bibinfo{author}{Bassett, D.~S.} (\bibinfo{year}{2016}{\natexlab{a}}).
\newblock \bibinfo{title}{Small-world propensity and weighted brain networks}.
\newblock {\it \bibinfo{journal}{Scientific reports}\/},  {\it
  \bibinfo{volume}{6}\/}.
\bibitem[{Muldoon et~al.(2016{\natexlab{b}})Muldoon, Pasqualetti, Gu, Cieslak,
  Grafton, Vettel \& Bassett}]{muldoon2016stimulation}
\bibinfo{author}{Muldoon, S.~F.}, \bibinfo{author}{Pasqualetti, F.},
  \bibinfo{author}{Gu, S.}, \bibinfo{author}{Cieslak, M.},
  \bibinfo{author}{Grafton, S.~T.}, \bibinfo{author}{Vettel, J.~M.}, \&
  \bibinfo{author}{Bassett, D.~S.} (\bibinfo{year}{2016}{\natexlab{b}}).
\newblock \bibinfo{title}{Stimulation-based control of dynamic brain networks}.
\newblock {\it \bibinfo{journal}{arXiv preprint arXiv:1601.00987}\/}, .
\bibitem[{M{\"u}ller \& Schuppert(2011)}]{muller2011few}
\bibinfo{author}{M{\"u}ller, F.-J.}, \& \bibinfo{author}{Schuppert, A.}
  (\bibinfo{year}{2011}).
\newblock \bibinfo{title}{Few inputs can reprogram biological networks}.
\newblock {\it \bibinfo{journal}{Nature}\/},  {\it \bibinfo{volume}{478}\/},
  \bibinfo{pages}{E4--E4}.
\bibitem[{Nenadovic et~al.(2008)Nenadovic, Hutchison, Dominguez, Otsubo, Gray,
  Sharma, Belkas \& Perez~Velazquez}]{Nenadovic2008}
\bibinfo{author}{Nenadovic, V.}, \bibinfo{author}{Hutchison, J.~S.},
  \bibinfo{author}{Dominguez, L.~G.}, \bibinfo{author}{Otsubo, H.},
  \bibinfo{author}{Gray, M.~P.}, \bibinfo{author}{Sharma, R.},
  \bibinfo{author}{Belkas, J.}, \& \bibinfo{author}{Perez~Velazquez, J.~L.}
  (\bibinfo{year}{2008}).
\newblock \bibinfo{title}{Fluctuations in cortical synchronization in pediatric
  traumatic brain injury}.
\newblock {\it \bibinfo{journal}{J Neurotrauma}\/},  {\it
  \bibinfo{volume}{25}\/}, \bibinfo{pages}{615--627}.
\bibitem[{Niven \& Laughlin(2008)}]{Niven2008}
\bibinfo{author}{Niven, J.~E.}, \& \bibinfo{author}{Laughlin, S.~B.}
  (\bibinfo{year}{2008}).
\newblock \bibinfo{title}{Energy limitation as a selective pressure on the
  evolution of sensory systems}.
\newblock {\it \bibinfo{journal}{J Exp Biol}\/},  {\it
  \bibinfo{volume}{211}\/}, \bibinfo{pages}{1792--1804}.
\bibitem[{Nudo(2006)}]{Nudo2006}
\bibinfo{author}{Nudo, R.~J.} (\bibinfo{year}{2006}).
\newblock \bibinfo{title}{Mechanisms for recovery of motor function following
  cortical damage}.
\newblock {\it \bibinfo{journal}{Curr Opin Neurobiol}\/},  {\it
  \bibinfo{volume}{16}\/}, \bibinfo{pages}{638--644}.
\bibitem[{O'Reilly et~al.(2013)O'Reilly, Croxson, Jbabdi, Sallet, Noonan, Mars,
  Browning, Wilson, Mitchell, Miller et~al.}]{o2013causal}
\bibinfo{author}{O'Reilly, J.~X.}, \bibinfo{author}{Croxson, P.~L.},
  \bibinfo{author}{Jbabdi, S.}, \bibinfo{author}{Sallet, J.},
  \bibinfo{author}{Noonan, M.~P.}, \bibinfo{author}{Mars, R.~B.},
  \bibinfo{author}{Browning, P.~G.}, \bibinfo{author}{Wilson, C.~R.},
  \bibinfo{author}{Mitchell, A.~S.}, \bibinfo{author}{Miller, K.~L.} et~al.
  (\bibinfo{year}{2013}).
\newblock \bibinfo{title}{Causal effect of disconnection lesions on
  interhemispheric functional connectivity in rhesus monkeys}.
\newblock {\it \bibinfo{journal}{Proceedings of the National Academy of
  Sciences}\/},  {\it \bibinfo{volume}{110}\/}, \bibinfo{pages}{13982--13987}.
\bibitem[{Pasqualetti et~al.(2014)Pasqualetti, Zampieri \&
  Bullo}]{pasqualetti2014controllability}
\bibinfo{author}{Pasqualetti, F.}, \bibinfo{author}{Zampieri, S.}, \&
  \bibinfo{author}{Bullo, F.} (\bibinfo{year}{2014}).
\newblock \bibinfo{title}{Controllability metrics, limitations and algorithms
  for complex networks}.
\newblock {\it \bibinfo{journal}{Control of Network Systems, IEEE Transactions
  on}\/},  {\it \bibinfo{volume}{1}\/}, \bibinfo{pages}{40--52}.
\bibitem[{Pestilli et~al.(2014)Pestilli, Yeatman, Rokem, Kay \&
  Wandell}]{pestilli2014evaluation}
\bibinfo{author}{Pestilli, F.}, \bibinfo{author}{Yeatman, J.~D.},
  \bibinfo{author}{Rokem, A.}, \bibinfo{author}{Kay, K.~N.}, \&
  \bibinfo{author}{Wandell, B.~A.} (\bibinfo{year}{2014}).
\newblock \bibinfo{title}{Evaluation and statistical inference for human
  connectomes}.
\newblock {\it \bibinfo{journal}{Nature methods}\/},  {\it
  \bibinfo{volume}{11}\/}, \bibinfo{pages}{1058--1063}.
\bibitem[{Posner \& Petersen(1989)}]{posner1989attention}
\bibinfo{author}{Posner, M.~I.}, \& \bibinfo{author}{Petersen, S.~E.}
  (\bibinfo{year}{1989}).
\newblock {\it \bibinfo{title}{The attention system of the human brain}\/}.
\newblock \bibinfo{type}{Technical Report} DTIC Document.
\bibitem[{Power et~al.(2011)Power, Cohen, Nelson, Wig, Barnes, Church, Vogel,
  Laumann, Miezin, Schlaggar \& Petersen}]{Power2011}
\bibinfo{author}{Power, J.~D.}, \bibinfo{author}{Cohen, A.~L.},
  \bibinfo{author}{Nelson, S.~M.}, \bibinfo{author}{Wig, G.~S.},
  \bibinfo{author}{Barnes, K.~A.}, \bibinfo{author}{Church, J.~A.},
  \bibinfo{author}{Vogel, A.~C.}, \bibinfo{author}{Laumann, T.~O.},
  \bibinfo{author}{Miezin, F.~M.}, \bibinfo{author}{Schlaggar, B.~L.}, \&
  \bibinfo{author}{Petersen, S.~E.} (\bibinfo{year}{2011}).
\newblock \bibinfo{title}{Functional network organization of the human brain}.
\newblock {\it \bibinfo{journal}{Neuron}\/},  {\it \bibinfo{volume}{72}\/},
  \bibinfo{pages}{665--678}.
\bibitem[{Radhu et~al.(2012)Radhu, Daskalakis, Guglietti, Farzan, Barr,
  Arpin-Cribbie, Fitzgerald \& Ritvo}]{Radhu2012}
\bibinfo{author}{Radhu, N.}, \bibinfo{author}{Daskalakis, Z.~J.},
  \bibinfo{author}{Guglietti, C.~L.}, \bibinfo{author}{Farzan, F.},
  \bibinfo{author}{Barr, M.~S.}, \bibinfo{author}{Arpin-Cribbie, C.~A.},
  \bibinfo{author}{Fitzgerald, P.~B.}, \& \bibinfo{author}{Ritvo, P.}
  (\bibinfo{year}{2012}).
\newblock \bibinfo{title}{Cognitive behavioral therapy-related increases in
  cortical inhibition in problematic perfectionists}.
\newblock {\it \bibinfo{journal}{Brain Stimul}\/},  {\it
  \bibinfo{volume}{5}\/}, \bibinfo{pages}{44--54}.
\bibitem[{Raichle(2015)}]{Raichle2015}
\bibinfo{author}{Raichle, M.~E.} (\bibinfo{year}{2015}).
\newblock \bibinfo{title}{The brain's default mode network}.
\newblock {\it \bibinfo{journal}{Annu Rev Neurosci}\/},  {\it
  \bibinfo{volume}{38}\/}, \bibinfo{pages}{433--447}.
\bibitem[{Raichle et~al.(2001)Raichle, MacLeod, Snyder, Powers, Gusnard \&
  Shulman}]{Raichle2001}
\bibinfo{author}{Raichle, M.~E.}, \bibinfo{author}{MacLeod, A.~M.},
  \bibinfo{author}{Snyder, A.~Z.}, \bibinfo{author}{Powers, W.~J.},
  \bibinfo{author}{Gusnard, D.~A.}, \& \bibinfo{author}{Shulman, G.~L.}
  (\bibinfo{year}{2001}).
\newblock \bibinfo{title}{A default mode of brain function}.
\newblock {\it \bibinfo{journal}{Proc Natl Acad Sci U S A}\/},  {\it
  \bibinfo{volume}{98}\/}, \bibinfo{pages}{676--682}.
\bibitem[{Raichle \& Snyder(2007)}]{Raichle2007}
\bibinfo{author}{Raichle, M.~E.}, \& \bibinfo{author}{Snyder, A.~Z.}
  (\bibinfo{year}{2007}).
\newblock \bibinfo{title}{A default mode of brain function: a brief history of
  an evolving idea}.
\newblock {\it \bibinfo{journal}{Neuroimage}\/},  {\it \bibinfo{volume}{37}\/},
  \bibinfo{pages}{1083--1090}.
\bibitem[{Raja~Beharelle et~al.(2012)Raja~Beharelle, Kovacevic, McIntosh \&
  Levine}]{Raja2012}
\bibinfo{author}{Raja~Beharelle, A.}, \bibinfo{author}{Kovacevic, N.},
  \bibinfo{author}{McIntosh, A.~R.}, \& \bibinfo{author}{Levine, B.}
  (\bibinfo{year}{2012}).
\newblock \bibinfo{title}{Brain signal variability relates to stability of
  behavior after recovery from diffuse brain injury}.
\newblock {\it \bibinfo{journal}{Neuroimage}\/},  {\it \bibinfo{volume}{60}\/},
  \bibinfo{pages}{1528--1537}.
\bibitem[{Rajan et~al.(2016)Rajan, Harvey \& Tank}]{Rajan2016}
\bibinfo{author}{Rajan, K.}, \bibinfo{author}{Harvey, C.~D.}, \&
  \bibinfo{author}{Tank, D.~W.} (\bibinfo{year}{2016}).
\newblock \bibinfo{title}{Recurrent network models of sequence generation and
  memory}.
\newblock {\it \bibinfo{journal}{Neuron}\/},  {\it \bibinfo{volume}{90}\/},
  \bibinfo{pages}{128--142}.
\bibitem[{Reveley et~al.(2015)Reveley, Seth, Pierpaoli, Silva, Yu, Saunders,
  Leopold \& Frank}]{reveley2015superficial}
\bibinfo{author}{Reveley, C.}, \bibinfo{author}{Seth, A.~K.},
  \bibinfo{author}{Pierpaoli, C.}, \bibinfo{author}{Silva, A.~C.},
  \bibinfo{author}{Yu, D.}, \bibinfo{author}{Saunders, R.~C.},
  \bibinfo{author}{Leopold, D.~A.}, \& \bibinfo{author}{Frank, Q.~Y.}
  (\bibinfo{year}{2015}).
\newblock \bibinfo{title}{Superficial white matter fiber systems impede
  detection of long-range cortical connections in diffusion mr tractography}.
\newblock {\it \bibinfo{journal}{Proceedings of the National Academy of
  Sciences}\/},  {\it \bibinfo{volume}{112}\/}, \bibinfo{pages}{E2820--E2828}.
\bibitem[{Rusconi et~al.(2009)Rusconi, Pinel, Eger, LeBihan, Thirion, Dehaene
  \& Kleinschmidt}]{rusconi2009disconnection}
\bibinfo{author}{Rusconi, E.}, \bibinfo{author}{Pinel, P.},
  \bibinfo{author}{Eger, E.}, \bibinfo{author}{LeBihan, D.},
  \bibinfo{author}{Thirion, B.}, \bibinfo{author}{Dehaene, S.}, \&
  \bibinfo{author}{Kleinschmidt, A.} (\bibinfo{year}{2009}).
\newblock \bibinfo{title}{A disconnection account of {G}erstmann syndrome:
  functional neuroanatomy evidence}.
\newblock {\it \bibinfo{journal}{Ann Neurol}\/},  {\it \bibinfo{volume}{66}\/},
  \bibinfo{pages}{654--662}.
\bibitem[{Rushworth et~al.(2006)Rushworth, Behrens \&
  Johansen-Berg}]{rushworth2006connection}
\bibinfo{author}{Rushworth, M.~F.}, \bibinfo{author}{Behrens, T.~E.}, \&
  \bibinfo{author}{Johansen-Berg, H.} (\bibinfo{year}{2006}).
\newblock \bibinfo{title}{Connection patterns distinguish 3 regions of human
  parietal cortex}.
\newblock {\it \bibinfo{journal}{Cereb Cortex}\/},  {\it
  \bibinfo{volume}{16}\/}, \bibinfo{pages}{1418--1430}.
\bibitem[{Salvador et~al.(2005)Salvador, Suckling, Schwarzbauer \&
  Bullmore}]{Salvador2005}
\bibinfo{author}{Salvador, R.}, \bibinfo{author}{Suckling, J.},
  \bibinfo{author}{Schwarzbauer, C.}, \& \bibinfo{author}{Bullmore, E.}
  (\bibinfo{year}{2005}).
\newblock \bibinfo{title}{Undirected graphs of frequency-dependent functional
  connectivity in whole brain networks}.
\newblock {\it \bibinfo{journal}{Philos Trans R Soc Lond B Biol Sci}\/},  {\it
  \bibinfo{volume}{360}\/}, \bibinfo{pages}{937--946}.
\bibitem[{Sharp et~al.(2014)Sharp, Scott \& Leech}]{Sharp2014}
\bibinfo{author}{Sharp, D.~J.}, \bibinfo{author}{Scott, G.}, \&
  \bibinfo{author}{Leech, R.} (\bibinfo{year}{2014}).
\newblock \bibinfo{title}{Network dysfunction after traumatic brain injury}.
\newblock {\it \bibinfo{journal}{Nat Rev Neurol}\/},  {\it
  \bibinfo{volume}{10}\/}, \bibinfo{pages}{156--166}.
\bibitem[{Shenoy et~al.(2011)Shenoy, Kaufman, Sahani \&
  Churchland}]{shenoy2011dynamical}
\bibinfo{author}{Shenoy, K.~V.}, \bibinfo{author}{Kaufman, M.~T.},
  \bibinfo{author}{Sahani, M.}, \& \bibinfo{author}{Churchland, M.~M.}
  (\bibinfo{year}{2011}).
\newblock \bibinfo{title}{A dynamical systems view of motor preparation:
  implications for neural prosthetic system design}.
\newblock {\it \bibinfo{journal}{Prog Brain Res}\/},  {\it
  \bibinfo{volume}{192}\/}, \bibinfo{pages}{33--58}.
\bibitem[{Sidaros et~al.(2008)Sidaros, Engberg, Sidaros, Liptrot, Herning,
  Petersen, Paulson, Jernigan \& Rostrup}]{Sidaros2008}
\bibinfo{author}{Sidaros, A.}, \bibinfo{author}{Engberg, A.~W.},
  \bibinfo{author}{Sidaros, K.}, \bibinfo{author}{Liptrot, M.~G.},
  \bibinfo{author}{Herning, M.}, \bibinfo{author}{Petersen, P.},
  \bibinfo{author}{Paulson, O.~B.}, \bibinfo{author}{Jernigan, T.~L.}, \&
  \bibinfo{author}{Rostrup, E.} (\bibinfo{year}{2008}).
\newblock \bibinfo{title}{Diffusion tensor imaging during recovery from severe
  traumatic brain injury and relation to clinical outcome: a longitudinal
  study}.
\bibitem[{Sizemore et~al.(2015)Sizemore, Giusti \&
  Bassett}]{sizemore2015classification}
\bibinfo{author}{Sizemore, A.}, \bibinfo{author}{Giusti, C.}, \&
  \bibinfo{author}{Bassett, D.} (\bibinfo{year}{2015}).
\newblock \bibinfo{title}{Classification of weighted networks through mesoscale
  homological features}.
\newblock {\it \bibinfo{journal}{arXiv preprint arXiv:1512.06457}\/}, .
\bibitem[{Sommer et~al.(2012)Sommer, Slotema, Daskalakis, Derks, Blom \&
  van~der Gaag}]{Sommer2012}
\bibinfo{author}{Sommer, I.~E.}, \bibinfo{author}{Slotema, C.~W.},
  \bibinfo{author}{Daskalakis, Z.~J.}, \bibinfo{author}{Derks, E.~M.},
  \bibinfo{author}{Blom, J.~D.}, \& \bibinfo{author}{van~der Gaag, M.}
  (\bibinfo{year}{2012}).
\newblock \bibinfo{title}{The treatment of hallucinations in schizophrenia
  spectrum disorders}.
\newblock {\it \bibinfo{journal}{Schizophr Bull}\/},  {\it
  \bibinfo{volume}{38}\/}, \bibinfo{pages}{704--714}.
\bibitem[{Sridharan et~al.(2008)Sridharan, Levitin \&
  Menon}]{sridharan2008critical}
\bibinfo{author}{Sridharan, D.}, \bibinfo{author}{Levitin, D.~J.}, \&
  \bibinfo{author}{Menon, V.} (\bibinfo{year}{2008}).
\newblock \bibinfo{title}{A critical role for the right fronto-insular cortex
  in switching between central-executive and default-mode networks}.
\newblock {\it \bibinfo{journal}{Proc Natl Acad Sci U S A}\/},  {\it
  \bibinfo{volume}{105}\/}, \bibinfo{pages}{12569--125674}.
\bibitem[{Stam(2014)}]{Stam2014}
\bibinfo{author}{Stam, C.~J.} (\bibinfo{year}{2014}).
\newblock \bibinfo{title}{Modern network science of neurological disorders}.
\newblock {\it \bibinfo{journal}{Nat Rev Neurosci}\/},  {\it
  \bibinfo{volume}{15}\/}, \bibinfo{pages}{683--695}.
\bibitem[{Stephan et~al.(2009)Stephan, Friston \&
  Frith}]{stephan2009dysconnection}
\bibinfo{author}{Stephan, K.~E.}, \bibinfo{author}{Friston, K.~J.}, \&
  \bibinfo{author}{Frith, C.~D.} (\bibinfo{year}{2009}).
\newblock \bibinfo{title}{Dysconnection in schizophrenia: from abnormal
  synaptic plasticity to failures of self-monitoring}.
\newblock {\it \bibinfo{journal}{Schizophr Bull}\/},  {\it
  \bibinfo{volume}{35}\/}, \bibinfo{pages}{509--527}.
\bibitem[{Szameitat et~al.(2011)Szameitat, Schubert \&
  M{\"u}ller}]{szameitat2011test}
\bibinfo{author}{Szameitat, A.~J.}, \bibinfo{author}{Schubert, T.}, \&
  \bibinfo{author}{M{\"u}ller, H.~J.} (\bibinfo{year}{2011}).
\newblock \bibinfo{title}{How to test for dual-task-specific effects in brain
  imaging studies—an evaluation of potential analysis methods}.
\newblock {\it \bibinfo{journal}{Neuroimage}\/},  {\it \bibinfo{volume}{54}\/},
  \bibinfo{pages}{1765--1773}.
\bibitem[{Thomas et~al.(2014)Thomas, Frank, Irfanoglu, Modi, Saleem, Leopold \&
  Pierpaoli}]{thomas2014anatomical}
\bibinfo{author}{Thomas, C.}, \bibinfo{author}{Frank, Q.~Y.},
  \bibinfo{author}{Irfanoglu, M.~O.}, \bibinfo{author}{Modi, P.},
  \bibinfo{author}{Saleem, K.~S.}, \bibinfo{author}{Leopold, D.~A.}, \&
  \bibinfo{author}{Pierpaoli, C.} (\bibinfo{year}{2014}).
\newblock \bibinfo{title}{Anatomical accuracy of brain connections derived from
  diffusion mri tractography is inherently limited}.
\newblock {\it \bibinfo{journal}{Proceedings of the National Academy of
  Sciences}\/},  {\it \bibinfo{volume}{111}\/}, \bibinfo{pages}{16574--16579}.
\bibitem[{Tortella et~al.(2014)Tortella, Selingardi, Moreno, Veronezi \&
  Brunoni}]{Tortella2014}
\bibinfo{author}{Tortella, G.}, \bibinfo{author}{Selingardi, P.~M.},
  \bibinfo{author}{Moreno, M.~L.}, \bibinfo{author}{Veronezi, B.~P.}, \&
  \bibinfo{author}{Brunoni, A.~R.} (\bibinfo{year}{2014}).
\newblock \bibinfo{title}{Does non-invasive brain stimulation improve cognition
  in major depressive disorder? a systematic review}.
\newblock {\it \bibinfo{journal}{CNS Neurol Disord Drug Targets}\/},  {\it
  \bibinfo{volume}{13}\/}, \bibinfo{pages}{1759--1769}.
\bibitem[{Weiss et~al.(2011)Weiss, Bassett, Rubinstein, Holroyd, Apud,
  Dickinson \& Coppola}]{weiss2011functional}
\bibinfo{author}{Weiss, S.~A.}, \bibinfo{author}{Bassett, D.~S.},
  \bibinfo{author}{Rubinstein, D.}, \bibinfo{author}{Holroyd, T.},
  \bibinfo{author}{Apud, J.}, \bibinfo{author}{Dickinson, D.}, \&
  \bibinfo{author}{Coppola, R.} (\bibinfo{year}{2011}).
\newblock \bibinfo{title}{Functional brain network characterization and
  adaptivity during task practice in healthy volunteers and people with
  schizophrenia}.
\newblock {\it \bibinfo{journal}{Frontiers in human neuroscience}\/},  {\it
  \bibinfo{volume}{5}\/}, \bibinfo{pages}{81}.
\bibitem[{Yan et~al.(2012)Yan, Ren, Lai, Lai \& Li}]{yan2012controlling}
\bibinfo{author}{Yan, G.}, \bibinfo{author}{Ren, J.}, \bibinfo{author}{Lai,
  Y.-C.}, \bibinfo{author}{Lai, C.-H.}, \& \bibinfo{author}{Li, B.}
  (\bibinfo{year}{2012}).
\newblock \bibinfo{title}{Controlling complex networks: How much energy is
  needed?}
\newblock {\it \bibinfo{journal}{Physical review letters}\/},  {\it
  \bibinfo{volume}{108}\/}, \bibinfo{pages}{218703}.
\bibitem[{Yeo et~al.(2011)Yeo, Krienen, Sepulcre, Sabuncu, Lashkari,
  Hollinshead, Roffman, Smoller, Zollei, Polimeni, Fischl, Liu \&
  Buckner}]{Yeo2011}
\bibinfo{author}{Yeo, B.~T.}, \bibinfo{author}{Krienen, F.~M.},
  \bibinfo{author}{Sepulcre, J.}, \bibinfo{author}{Sabuncu, M.~R.},
  \bibinfo{author}{Lashkari, D.}, \bibinfo{author}{Hollinshead, M.},
  \bibinfo{author}{Roffman, J.~L.}, \bibinfo{author}{Smoller, J.~W.},
  \bibinfo{author}{Zollei, L.}, \bibinfo{author}{Polimeni, J.~R.},
  \bibinfo{author}{Fischl, B.}, \bibinfo{author}{Liu, H.}, \&
  \bibinfo{author}{Buckner, R.~L.} (\bibinfo{year}{2011}).
\newblock \bibinfo{title}{The organization of the human cerebral cortex
  estimated by intrinsic functional connectivity}.
\newblock {\it \bibinfo{journal}{J Neurophysiol}\/},  {\it
  \bibinfo{volume}{106}\/}, \bibinfo{pages}{1125--1165}.

\end{thebibliography}

\end{document}